\documentclass[12pt, preprint]{aastex}
\usepackage{emulateapj5}  %  ApJ journal style

\def\einstein{{\it Einstein }}	% ApJ uses Italics for 'Einstein' Satellite 
\def\ergcm2s{~erg cm$^{-2}$ s$^{-1}$} % ergs per cm**2 second 
\def\ergs{~erg s$^{-1}$}		% ergs per second 
\def\lunit{~erg s$^{-1}$}		% ergs per second 
\def\cmcube{~cm$^{-3}$}		% 
\def\cmsq{~cm$^{-2}$}		% 
\def\nh{~$\rm{N_{H}}$}

\def\etal{et al.~}		% Present ApJ style, override older definitions
\def\kmsmpc{~km s$^{-1}$ Mpc$^{-1}$}		% kilometers per second 
\def\msun{~M$_{\odot}$ }
\def\n4038{~NGC4038/39}		% Use in text.  For title, use NGC 4038/4039
\def\chandra{{\it Chandra }}
\def\x2{$\chi^{2}$}	

\begin{document}

\title{The X-ray Source Population of ``The Antennae'' Galaxies:
 X-ray Properties from {\textit{CHANDRA}} and  Multiwavelength Associations}

\author{ A. Zezas, G. Fabbiano, A. H. Rots, S. S. Murray}
\affil{Harvard-Smithsonian Center for Astrophysics,\\ 60 Garden
Street, Cambridge, MA 02138}
\shorttitle{\chandra\ The X-ray Source Population of  ``The Antennae'' Galaxies }
\shortauthors{Zezas et al.}
\bigskip

\begin{abstract}
We  investigate the nature  of the luminous X-ray source population 
detected in a (72~ks) \chandra ACIS-S observation of NGC~4038/39, the Antennae galaxies.
We derive the average X-ray spectral
properties of sources in different luminosity ranges, and
we correlate the X-ray positions with radio, IR, and optical ({\it HST})
data. The X-ray sources are predominantly associated with young stellar
clusters,  indicating that they belong to the young stellar
population.
Based on both their co-added X-ray spectrum, and on the lack of associated radio emission,
we conclude that the Ultra Luminous X-ray sources (ULXs),
with $L_X \ge 10^{39} \rm ergs~s^{-1}$, are not young compact Supernova Remnants
(SNR), but accretion binaries. While their spectrum  is consistent with those of
ULXs studied in nearby galaxies, and interpreted
as the counterparts of intermediate mass black-holes (M$>10-1000$\msun),
comparison with the position of star-clusters suggests that some
of the ULXs may be runaway binaries, thus suggesting lower-mass binary systems.
The co-added spectrum of the sources in the
$3\times10^{38}-10^{39}$\ergs~ luminosity range is consistent with
those of Galactic black-hole candidates. These sources are also on average
displaced from neighboring star clusters.
The softer spectrum of the less luminous sources suggests 
the presence of SNRs or of hot interstellar 
medium (ISM) in the {\it Chandra} source extraction area.
Comparison with HI and CO observations shows that most sources are detected
in the outskirts of large concentrations of gas. The absorbing columns 
inferred from these observations would indeed absorb X-rays up to 5~keV,
so there may be several hidden X-ray sources. Associated with
these obscured regions we find 6 sources with heavily
absorbed X-ray spectra and absorption-corrected luminosities in the ULX
range. We detect the nuclei of both galaxies with luminosities in the $10^{39} \rm ergs~s^{-1}$
range and soft, possibly thermal, X-ray spectra.

\end{abstract}


\section{Introduction}

X-ray studies of normal and actively star-forming galaxies with the \einstein
Observatory and {\it ROSAT} suggested that their X-ray emission
arises from a population of discrete sources (X-ray binaries,
XRBs, and supernova remnants,
SNRs), as well as from
superwinds and hot ISM associated with starburst activity  (for
reviews see Fabbiano 1989, 1995).
A few of the point-like sources were found to have extremely  high
X-ray luminosities, when compared to the analogous populations of
the Milky Way or M31 (e.g., Fabbiano 1995). These luminosities are well
in excess of the Eddington luminosity for an accreting
neutron star and have led to speculations that the Ultra-Luminous X-ray
sources (ULXs) may be associated with intermediate mass black holes 
(10-100~M$_{\odot}$;
e.g., Fabbiano 1989, Zezas \etal 1999, Roberts \& Warwick 2000, Makishima \etal 2000).

Although these studies suggested both variety and complexity in the X-ray
source populations of galaxies, beyond what can be observed in our
local environment, it is only now with the sub-arcsecond resolution of
\chandra (Van Speybroeck \etal 1997; Weisskopf \etal 2000) that studies of galactic
 X-ray source populations can be attempted at large.
With \chandra individual X-ray sources
are not likely to be confused in observations of galaxies as far
as the Virgo cluster and beyond. Moreover, the high spatial resolution of
\chandra  results in very low background contamination, allowing the
detection of very faint sources.

In this paper we discuss the nature of 49 luminous X-ray 
sources detected with \chandra in the actively star-forming
merging galaxies NGC~4038/39 (\chandra Observation ID 315; see Fabbiano,
Zezas \& Murray 2001 for a first overall discussion of these data), by
means of their average X-ray spectral properties and possible 
conterparts at different wavelengths. The detection of these
sources and their detailed X-ray analysis is reported in Zezas et al (2002; Paper II).
This population of X-ray sources is exceptional, because of the 
very large number of ULXs in a single galaxy.

 The Antennae are the prototypical system of merging galaxies
(Toomre \& Toomre 1977), and a wealth of multi-frequency 
observational data are available. 
The system is in an early merging phase about $3\times10^{8}$~yr  before
full coalescence (Georgakakis \etal 2001), when
intense bursts of star-formation are likely to occur (e.g. Mihos \& Hernquist 1996).
Optical and near-IR observations show that there are
several generations of stellar populations with ages ranging from 6 Myr
to $\sim 1$~Gyr (e.g. Whitmore \etal 1999,
Kunze \etal 1996, Mengel \etal 2001, Fisher \etal 1996).
In the mid-IR (10-20$~\mu$m), as well as in molecular CO, 
the dust-enshrouded contact region 
of the two merging galaxies, where most of the
star formation is presently occurring, dominates the emission 
(Mirabel \etal 1998, Wilson \etal 2001).
A large number of both non-thermal (SNRs) and thermal
(HII regions) radio sources are detected in VLA observations at 6~cm and 20~cm
(Neff \& Ulvestad 2000).  Zhang \etal (2001) performed a
multiwavelength study of the young star-clusters in the Antennae
galaxies. They find that the youngest clusters (younger than 5~Myr)
are associated preferentially with radio continnum, HI and CO emission
whereas slightly older clusters (ages 5 - 160~Myr) are associated with
H$\alpha$, UV and X-ray emission (the latter based on ROSAT data). 

This paper is organized as follows: in \S2. we summarize the
results of Paper~II that are pertinent to the present discussion;
in \S3. we further investigate the properties of the X-ray
source population, by analyzing the average
spectral properties in different luminosity ranges;
in \S4. we obtain additional contraints to the nature and
evolution of the X-ray sources, by examining their 
multi-wavelength environment; in \S5 we discuss our result;
our conclusions are summarized in \S6.
Throughout this paper we use a distance of 29~Mpc ($\rm{H_{0}=50}$\kmsmpc).
For $\rm{H_{0}=75}$\kmsmpc, the distance becomes 19.3~Mpc  
and all the cited luminosities will be a factor of 2.25 lower. At a
distance of 29~Mpc, 1~arcsec corresponds to a physical distance of
$140$~pc (at 19~Mpc it corresponds to 92~pc). All cited luminosities
are in the 0.1-10.0~keV band, unless otherwise stated.

\section{Summary of Individual X-ray Source Properties}

We refer to  Paper II (Zezas \etal 2002) for a detailed analysis of
the discrete X-ray sources discovered with \chandra in the Antennae.
Here we summarize the main results:

\medskip
\noindent
1) Forty-nine sources are detected down to a
detection limit of $\sim10^{38}$~\ergs ($\rm{H_{0}=50}$\kmsmpc;
$5\times10^{37}$~\ergs for $\rm{H_{0}=75}$\kmsmpc). 
A soft (0.3 - 2.0~keV) and a medium (2.0-
4.0~keV) band image of the Antennae is presented in Figure~1. The
 detected sources are marked by the 3$\sigma$ ellipses fit to the
spatial distribution of the source events, and numbered following the
numbering convention of Table~1 in Paper~II.
Only 6 of these
sources, including the Southern nucleus and one source in the Northern
nuclear region, are associated with small-scale 
extended X-ray emission (sources 5, 6, 7, 10, 24, 29; identified with
light blue ellipses in Fig.~1); the other 43 sources appear point-like. 

\smallskip
\noindent
2) Of the 49 sources, 31  have
luminosities below $\sim10^{39}$~\ergs ($\rm{H_{0}=50}$\kmsmpc), while
18 are ULXs with absorption corrected luminosities ranging from $10^{39}$~\ergs up to
$2\times10^{40}$~\ergs, for a 5~keV Bremsstrahlung model and Galactic
line-of-sight \nh.

\smallskip
\noindent
3) There is some evidence of time variability in the ULXs:
two sources (14, 44) are found to vary during the \chandra
observation and 3 of the 6 sources for which 
 a comparison with the 5" resolution
{\it ROSAT HRI} observations (Fabbiano \etal 1997) is 
possible, were found to be variable (16, 42, 44/46) in timescales of years. 
These variable sources are identified by green ellipses
in Fig.~1.

\smallskip
\noindent
4) We fitted the data with both an absorbed power-law model and a thermal Raymond-Smith 
model with solar abundances and absorption.
In most cases the power-law model favors column densities in excess of the
Galactic line of sight \nh, while the Raymond-Smith model requires
much lower absorbing column densities, without significantly
improving the fit.
Two-component models (power-law plus
Raymond Smith) were fitted to
the spectra of 10 sources for which a large enough number of counts was detected. In all 
these fits, the overall $N_H$ is consistent with the Galactic value,
while the hard component has usually a higher absorption.  Three
sources show significant fit residuals suggesting the presence  
of emission lines.

\smallskip
\noindent
5) Six sources  have large best-fit column densities  (greater than 10
times the line of sight Galactic $N_H$) in both the power-law and the
double component spectral fits, suggesting truly
absorbed spectra. These sources (12, 24, 25, 34, 35, 36) are identifyed by black ellipses
in Fig.~1, and reside in or near the very dusty contact region of the
two merging galaxies.

\smallskip
\noindent
6) There is a suggestion both from the X-ray colors (Hardness Ratios) and 
the power-law slopes in the simple power-law fits, that more luminous
sources tend to be harder. 
Photon indices range from $\Gamma\sim 1.2$ for the most luminous ULXs to
 $\Gamma > 3.0$ for fainter sources, but the uncertainties are very large in most cases,
because of the limited statistics.

\section{Co-added X-ray Spectra}

Detailed observations of luminous X-ray sources in the Milky Way
and Magellanic Clouds, augmented more recently by a few observations
in nearby spirals galaxies, have produced a `library' of spectral
shapes characteristics of different types of sources, which in the 
case of accreting binaries may also depend on the intensity 
of the source at a given time.
Pulsar XRBs as well as Black Hole (BH) binaries in  low state tend to have hard spectra
($\rm{\Gamma<2.0}$;  e.g. Nagase 1989, Tanaka \& Lewin 1995). The spectra of BH binaries in  high state
 and unmagnetised neutron stars are dominated by a multi-temperature
disk black-body (disk-BB)  spectrum (e.g. van Paradijs 1999,
 Tanaka \& Lewin 1995).  ULXs in nearby galaxies have also been
found to oscillate between these intensity/spectral states (e.g. 
La Parola \etal 2001; Kubota \etal 2001). SNRs  typically exhibit soft thermal emission
(kT~$\sim0.5-5.0$~keV; e.g., Schlegel 1995), although young supernovae detonating in dense
environments (cSNRs) can have much harder spectra (Plewa 1995, Terlevich 1994).

The general spectral trend found in Paper~II for the 
sources in the Antennae suggests that the ULXs may resemble
similarly hard ULXs found in other galaxies, which on
the basis of their spectral behaviour were identified with accretion binaries (e.g.
LaParola et al 2001). However,  even for these sources the number of detected
photons is not high enough to pursue a detailed spectral investigation.
The situation is even harder for fainter sources.

For this reason we investigated the average spectral properties
of sources in three different luminosity ranges,
by extracting and analyzing co-added spectra for the sources in each
group.  The underlying assumption is that, in each luminosity
range, the source population may be dominated by a given type of sources,
so that the coadded spectrum may express the average spectral characteristics
of these sources. While the conclusive work will require future deeper 
observations, this assumption is supported by the general characteristics
of the Galactic  X-ray sources (e.g. Watson 1990) and by the general luminosity-spectral 
trend reported in Paper~II. 

In {\it Group 1} we included sources with $L_X < 3\times10^{38}$\ergs~ which is the
Eddington luminosity for the upper range of neutron star masses. 
If $\rm{H_{0}=75}$~\kmsmpc~ this limit becomes $1.3\times10^{38}$~\ergs,
the Eddington luminosity for 1~\msun~ object.
{\it Group 2} includes sources with
luminosities between $3\times10^{38}$ and $10^{39}$\ergs~
($1.3\times10^{38} -4\times10^{38}$~\ergs~ for
$\rm{H_{0}=75}$~\kmsmpc), that may host stellar mass black holes, 
if the spherical accretion paradigm applies.
{\it Group 3} includes the ULXs. 

Co-adding spectra may introduce spurious
features, mainly below 1.0~keV,  due to the varying column density of
the individual spectra. However, the spectral parameters we are
interested in (the power-law photon index $\Gamma$, and the
temperature of the accretion disk) mostly depend on the
shape of the continuum above 1.0~keV. For the same reason, calibration
uncertainties below   1.0~keV  do not affect our conclusions.
In order to account for these uncertainties,
 we include 10\% systematic errors in the spectra of the brightest
sources,  in the region of the Oxygen edge ($\sim0.5$~keV). Because of 
variations of the intensity of the diffuse emission and of  the area
used to extract the source spectra, we cannot use the local background of each source,
since the background spectrum will not be correctly normalized to the
source area.  Therefore, the  background spectrum was
taken from a large, source-free area around the galaxy. For this reason,
the co-added spectra  may contain some
residual  local diffuse emission. We analyzed these spectra
following the procedures described in Paper~II.

The coadded  spectra, together with the best fit models and the
fit residuals, are presented in Fig.~2.  Because of the
higher S/N of the co-added spectra we were able to fit a wider range
of models to the data: single component power-law (PO) and optically thin
thermal Raymond-Smith models (RS), a combination of PO and RS (PO+RS),
a PO and a multi-temperature disk black-body (PO + disk-BB), a PO+disk-BB with a
thermal RS component (PO+RS+disk-BB) and finally a PO+RS+RS model.
  The results of the spectral fitting are
given in Table~1A.  The first column gives the source luminosity range of the
co-added spectra, Columns (2) and (3) give the parameters of the PO
fits ($\Gamma$ and \nh~  in the first line and \x2 and dof in the second
line); Columns (4) and (5) give the parameters of the PO+RS model
($\Gamma$ and kT in the first line and \x2 in the second line);
Columns (6) and (7) give the parameters of the PO + disk-BB
model ($\Gamma$ and the inner temperature of the disk, in the
first line and the \nh~ and \x2 (dof) in the second line); Columns (8)
and (9) give the parameters of the PO + RS + disk-BB model fits
($\Gamma$ and kT in the first line, inner temperature of the disk and
\nh~ in the second line, and \x2 in the third line); Columns (10)
and (11) give the parameters of the PO + RS + RS model ($\Gamma$ and
kT in the first line, kT and \nh~ in the second line and \x2 in the
third line). All temperatures are in units of ~keV and the \nh~ is
 in units of $10^{22}$\cmsq. The parameters of the best fit models are
 shown in boldface.

 Single component models gave an unacceptable fit in all cases. Fits
to composite   PO+RS models gave an  improved  fit
at more than 99\% confidence level for 2 additional parameters. This 
composite model gave a very good fit for the spectrum of Group 1,
but is still unacceptable for Groups 2 and 3. 
A  composite PO + disk-BB
model (that was found to represent well certain Galactic sources, and
ULXs in nearby spirals, e.g. Kubota \etal 2001), in the case of Group 1, 
gave a fit of similar quality to
the PO+RS model, but for Groups  2  and 3 the quality of the fit was still
unacceptable. PO photon indeces $\Gamma$ tend to be large for fainter
sources (Group~1) in agreement with the trends found in Paper~II.

Including a third component (PO+RS+RS or PO+RS + disk-BB model)
to fit the spectra of Groups 2 and 3, improved the fit in both cases.
The PO+RS + disk-BB model gave a slightly improved the fit for Group~3.
The best-fit power-laws are fairly hard, reaching values of $\Gamma \sim 1.5$
for the disk-BB model fits. In this model, the inner temperature of
the accretion disk is higher for the ULXs, reaching values comparable
to those reported by studies of individual ULX spectra with ASCA
(Makishima et al 2000).

The residuals from these best fit models for groups 1 and 2 showed
line-like features. We modeled these features by adding 
narrow gaussian lines to the models. Group 1 requires only one
additional line at 1.3~keV corresponding \ion{Mg}{9}.
Group 2 requires two lines, one at $\sim1.8$~keV
and another one at $\sim2.5$~keV corresponding to the \ion{Si}{10} - \ion{Si}{14}
triplet and \ion{S}{14}, respectively. 
We note that using a double thermal model (RS+RS) does not improve the fit
compared to both the PO+RS and the PO+RS+gauss models.

 The overall best fit models are the PO+RS for Group 1, the PO+RS+RS
(or PO+disk-BB+RS)
for Group 2 and the PO+RS+disk BB (or PO+RS+RS) for Group 3. In Table~1B we give the
 relative contribution of each component to the soft (0.1-2.5)~keV and hard
(2.5-10.0)~keV bands: Column (1) gives the luminosity range of the
co-added spectra, Column (2) gives the spectral component, Columns (3)
and (4) give the contribution of each component in the soft and hard
band (normalized by the total flux in each band).
The RS component contributes relatively more in the fainter X-ray sources,
in agreement with the spectral softening reported in Paper~II.
However, this effect could also be due to the relatively larger
contribution of the hot ISM of the Antennae (Paper~I) to the spectral
extraction areas of fainter point-like sources.

\section{Multiwavelength study}

The discrete source population of the Antennae tends to follow
the areas where the H$\alpha$ emission is more intense
(Paper~I), suggesting a connection with the young stellar
component. To better constrain the nature of these sources,
we have also looked at their  multi-wavelenth emission 
environment, using publicly available data and papers in
the literature. In particular, we have compared the X-ray 
positions with the star clusters discovered in the HST WFPC2
observations (Whitmore \etal 1995, 1999), to establish if our
sources can be associated with a star cluster; we have compared
our sources with the radio source lists of Neff \& Ulvestad (2000),
to establish if some of these may be associated with SNRs;
comparisons with both HST and IR data (e.g., Mengel et al 2001)
constrain the age of the associated stellar cluster, and therefore
the age and evolution of the X-ray source; comparisons between
the X-ray absorption columns and HI and CO observations help
to establish the 3-dimensional locations of these sources in the parent galaxies.
The results of these comparisons are given below.

\subsection{Optical}

The Antennae galaxies have been extensively observed with the WFPC and the WFPC2
on board HST (Whitmore \etal 1995, 1999). We obtained from the STScI
archive the WFPC2 observations 
in the F336W, F439W, and F814W  broad band filters (which correspond to
the U, B, I Johnson filters, respectively) as well as in the F658N filter which
corresponds to redshifted  H$\alpha$. 
These data were cleaned for
cosmic ray events and reduced following the standard procedures described in
the WFPC2 data analysis tutorial. Figure 3 shows the 3$\sigma$
ellipses to the
spatial distribution of the X-ray source events overlaid on  the U, H$\alpha$
and I band images. Because of the small field of view of WFPC2, 10
X-ray sources (4 of which are ULXs) are not included in this figure.
Yellow ellipses correspond to sources with X-ray luminosity above
$10^{39}$\ergs, green  ellipses correspond to sources
with X-ray luminosities between $3\times10^{38}$ and
$10^{39}$\ergs, and blue ellipses to
sources less luminous than $3\times10^{38}$\ergs. 
Marked with an ``X'' are  the 50 brightest young star-clusters (age $<30$Myr)
in the Antennae, from Table 1 of Whitmore
\etal (1999). Marked with an ``+'' the 25 brightest intermediate age
($\rm{0.25Gyr<T<1Gyr}$) and diamonds the globular 
clusters from Tables 2 and 3 of the same work. Foreground stars are
marked by circles. The ages of the
clusters are  based on their detection in the H$\alpha$ band and their
optical colours, and are also obtained from Whitmore \etal (1999).
Unfortunately, because of the small field of view of  the WFPC2 it is
not possible to identify sources common to the optical  and
the X-ray images, which we could use to compare  their absolute
astrometry.
 Therefore, we assume that
the absolute astrometry of the WFPC2 images is good within $0.9''$ to $1.5''$
arcsecond (Biretta \etal 2000).

 The properties of the optical counterparts of the 39 X-ray sources
within the WFPC2 field, are
listed in Table 2A, where column (1) gives the number of the X-ray
source from Table 1 in Paper~II, Column (2) gives its photon index and Columns (3) and (4) give
Logarithm of the soft (0.1-2.5~keV) and hard band (2.5-10.0~keV) luminosities
respectively, corrected for Galactic absorption, in units of \lunit. Column (5) gives the optical source number
from Tables 1 and 2 of Whitmore \etal (1999). Column (6) gives
the angular distance of the X-ray source from
each optical counterpart and Columns (7)-(10) give the V band
absolute magnitude, and the U-B, B-V, V-I colors
respectively. Finally Column (11) gives the evolutionary state of each cluster
based on the optical colors.  
 We consider an X-ray source as 
coincident with an optical source if their separation is less that
$2''$. This takes into account the uncertainties in the absolute
astrometry of \chandra and HST. Optical sources which fall
within the $3\sigma$ ellipse of an X-ray source, but have offsets
larger than 2$''$, are considered as
possibly associated. We find 8 X-ray sources coincident with 18 young
stellar clusters (separation smaller than $2''$) and 13 sources
possibly associated with 25 young and 2 intermediate age stellar
clusters. These numbers are only indicative since the list of optical
clusters we used in not complete by any means. There may well be more
sources with fainter optical counterparts as is discussed later.

In order to quantitatively assess
the chance coincidence probability, we created 10 source lists from
the original optical source lists but with the coordinates of each
source shifted by a random amount ranging between $\pm5''$ and
$\pm25''$ in both RA and Dec. These limits were chosen in order to
make sure that the new random sources will neither fall inside the
positional error box of the original sources nor will fall outside the
region of maximum source density in the galaxy. The latter is
particularly important since the distribution of both the optical and
X-ray  sources is not uniform (for the same reason we cannot simply
perform a  calculation based on the surface density of the sources in
each band). We find that $6\pm2$ X-ray sources could be associated by
chance with $8\pm4$ optical sources. So the majority of the associations 
are likely to have a real astrophysical origin. 

We searched for fainter
optical counterparts using the star clusters reported in the earlier
work of Whitmore \& Schweizer (1995), who present the full list of
star-clusters detected in an HST/WFPC observation of the Antennae down
to a limiting V magnitude of  $\sim23$. However, since these data were
obtained with WFPC before the HST mirror refurbishing 
 the spatial resolution and the  positional accuracy of
the sources is not as good  as for the later WFPC2 data. We find a
total of 25 X-ray sources with 177 optical counterparts from the
Table~1 of Whitmore \& Schweizer (1995). In Table 2B we present the
properties of the secure optical counterparts (offset less than $2''$):
Column (1) gives the X-ray source, Column (2) gives the optical source
from Table~1 of Whitmore \& Schweizer (1995), Column (3) gives the
distance between the optical and the X-ray counterpart, Column (4)
gives the V band magnitude of each cluster and Column (5) gives the
V-I colour (also from  Whitmore \& Schweizer 1995).
We find that, in addition to the 8 X-ray sources with secure bright optical
counterparts,  there are 16 X-ray sources with faint optical
counterparts whithin a $2''$ error circle.  Of the 14 ULXs in the
WFPC2 field, only 10
have a secure optical counterpart, while 4 have counterparts with
larger offsets ($2''-4''$).  For the 25 less luminous X-ray sources,
encompassed by the
WFPC2 field, 14 have a secure optical counterpart. The other 11 may
either have a fainter counterpart or a larger offset.

In Fig. 3d we show the offsets of each source from its nearest optical
counterpart for the ULXs (upper left) and sources in the
$3\times10^{38}-10^{39}$ luminosity range (upper right). For this figure we
used the source lists of Whitmore \etal (1999) as well as the more
complete lists of Whitmore \& Schweizer (1995).  The random
distribution of these offsets suggests that they are not due to a
misalignment of the X-ray and optical images. From the inspection of
the WFPC2 images we find that some X-ray sources have optical
counterparts which are closer than those tabulated in Whitmore \etal
(1999)  or Whitmore \& Schweizer (1995).  Therefore, in order to
assess the reality of the measured offsets we detected all the sources
in the I-band WFPC2 image using the IRAF {\textit{daofind}} task
 as described in Whitmore \& Heyer (1995).  We set a detection limit
of  7$\sigma$  above the local background of each chip, in order
to minimize the detection of stars and spurious sources. We find a
total of $\sim2000$ sources over the WFPC2 field of view. The bottom
panel of Fig. 3d shows the offsets between the X-ray sources and their nearest
optical source from this list (ULXs left and sources in the
$3\times10^{38}-10^{39}$ luminosity range in the right). It is clear that the inclusion of
fainter sources does not alter the fact that some X-ray sources are
offset even up to $\sim2''$ from their nearest optical source. Whitmore \etal (2002, private
communication) indicated that there is an offset of $\sim1''$ in the
absolute astrometry of WFPC2. However, the random distribution of the offsets
we measure, suggests that they are not
due to discrepancies between the absolute astrometry of the optical
and \chandra observations.  

%Chance coincidence for whitmore 95 is 24.4 \pm4.5 X-ray sources and
%82.6 \pm 21 optical srcs

\subsection{Radio}

The Antennae galaxies have recently been observed with the VLA in the
BnA, CnB and B configurations at 6cm and 4cm (Neff \& Ulvestad
2000; NU2000). The beam-size was $1.72''\times1.52''$ for the 6cm observations
and $1.36''\times0.96''$ for the 4cm observations. The beam area of
these observations  is $\sim10$ times larger than that of \chandra.
 In Figure 4 we show the soft (0.3-2.0~keV) and medium
(2.0-10.0~keV) band X-ray images with the radio sources marked as
crosses if they have a flat radio index ($\alpha>-0.4$) and X if
they have a steeper
index. On the same image we have plotted the X-ray source error
ellipses following the notation of Fig. 4.

 From these images it is clear  that there are some X-ray sources
positionally associated with radio sources. Again, we consider an X-ray and
radio source as coincident 
if their separation is
smaller than $2''$, and as possibly coincident if their separation is
larger but the radio source falls within the 
$3\sigma$ ellipse of the X-ray source.   We find a total of 22 X-ray
sources associated with 41 radio sources detected in either band; 16
of them have steep non-thermal radio spectra, 5 of which are ULXs. We  find 6
X-ray sources with a flat thermal spectrum radio counterpart, but none of them
is a ULX.   It
is possible that one X-ray source may coincide with more than one radio
source.  Table 3 presents the X-ray and  radio properties of these
sources: Columns (1) and (2) give the names of the X-ray and radio
sources. We assign two numbers to each radio source: the first is the
region number from the list of NU2000 and the second correspond to the
order of the source in each region (in Table 5 of NU2000). Columns (3)-(6)
give the distance from the X-ray source in arcseconds and the flux of the sources
detected in the 6~cm and 4~cm bands in $\mu$Jy. For the distances and the fluxes
we used the data from  Table 5 of NU2000 in order to have the same spatial
resolution in both bands. Columns (7)-(9) give the soft (0.1-2.5~keV)
and hard (2.5-10.0~keV) band luminosities, corrected for Galactic absorption, and the photon index for the
X-ray sources. Finally, Columns (10) and (11) give the radio luminosity
at 6~cm and the 4-6~cm radio spectral index.  In the case of radio sources
which belong to larger groups of sources we give the parameters of
both the individual sources and the group. Usually the radio spectral
index  refers to the integrated emission of the group, but is
dominated by the brightest sources.  In order to estimate the chance coincidence
probability we followed the same procedure as with the optical
sources. We estimate $11\pm3$ X-ray sources to be possibly coincident by chance
with $24\pm9$ radio sources.

\subsection{Infrared}

The Antennae galaxies have been extensively observed in the infrared
band. ISO observations (with a spatial resolution of $\sim20''$)
showed that there is a large amount of diffuse
 dust associated with the system (Mirabel \etal 1998, Vigroux \etal 1996). Low spatial resolution  mid-IR
spectroscopy with the ISO SWS and LWS showed that the IR emission of
this galaxy is dominated by early type stars (Fisher \etal 1996;
Kunze \etal 1996),  and give  a lower limit  of 65M$_{\odot}$ for
the upper mass cutoff of the IMF. The effective temperature of the
stellar populations  corresponds to that of O5 type
stars.
The IR spectrum of the obscured overlap region is well fit with a young starburst
($\sim7\times10^{6}$yr). 
 
More recently, Mengel  \etal  (2001) presented near-IR imaging/spectroscopy of 5
star-clusters detected in the H$\alpha$ band. Four of these clusters
contain X-ray sources (two of them are the nuclei) and one is in the
neighborhood of an X-ray source.  Stellar
population synthesis modeling of these star clusters (coincident with
our sources 34, 3 and 5) shows that their populations are very young
(ages $\sim3.7-8$Myr) with very large effective temperatures
($\sim35-40$K), consistent with the ISO results. Gilbert \etal (2000)
also presented NIR spectroscopy of the star-clusters associated with
source~3. They also found that it is very young ($\sim4$~Myr)
and highly obscured ($\rm{A_{V}=9-10}$~mag).  In contrast the two
nuclei have much older populations (ages $\sim65$~Myr). Although the
southern nucleus seems not to host any active star-formation (Gilbert
\etal 2000; Mengel \etal 2001), the northern nucleus (NGC 4038) has both
 an old population ($\sim60$~Myr) and a concentration of young stars
($\sim6$~Myr) located northward of the K-band peak.

\subsection{Atomic and molecular gas observations}

Recently Gordon \etal (2001) observed the Antennae galaxies in the
21cm HI line  with the Australia Telescope Compact Array. They find a total HI
mass of  $\sim5\times10^9$M$_{\odot}$,  about half of which
 is concentrated in the disk.  They also
found large concentrations of gas in the overlap region.  This large pool of gas  can
fuel vigorous bursts of star-formation as the merging evolves. 
 Fig. 5 presents an VLA HI image obtained by J. Hibbard (Hibbard
\etal 2001). The resolution of
this image is $11.4\arcsec\times7.4\arcsec$.  We have plotted contours in
different levels of column density (using the conversion factors from Hibbard
\etal 2001). Three things are clear from this image. a) The
southern nucleus is situated in a large reservoir of gas, whereas the
 atomic hydrogen in the area of the northern nucleus is depleted. b)
Most of the discrete X-ray sources are found in the outskirts of large
concentrations of gas. This could be a selection effect since the
 column density  in these regions is typically
$10^{23}-10^{24}$\cmsq, enough to absorb X-rays up to 5~keV or more. 
c) The fact that the measured column densities from the X-ray
spectra are much lower than those inferred by the HI image (typically one
order of magnitude less), suggests that the detected sources lie on
the surface of the galactic disks. Any sources embedded in the disks, or
on their far side, cannot be observed because of obscuration. 

The Antennae have also been observed in  CO  bands (Wilson \etal
2001). These observations showed that most of the molecular gas is in the
two nuclei and  5 supergiant molecular clouds. Although,
 the peak associated with the northern nucleus is not
positionaly coincident with the HI peak in the same region,  the
CO peaks in the five molecular clouds are spatially coincident with
the HI peaks.

\section{Discussion}

\subsection{Ultra Luminous X-ray sources: $\rm{L_{X}>10^{39}}$\ergs }

Because of their extreme luminosities, exceeding those
expected from a neutron star or stellar mass black-hole binary,
ULXs are intriguing sources found in a number of external galaxies
(e.g. Fabbiano 1995; Marston \etal, 1995). 
If they are accretion binaries, with spherically accreting
black holes, black hole masses may be in the $\rm{M\sim 10-1000s~M_{\sun}}$
range (e.g. Makishima \etal 2000,  Ptak \& Griffiths, 1999).
Other suggested possibilities include beamed lower-mass binaries
(King \etal 2001), and luminous 
SNRs exploding in dense media ($\sim10^{6}$~\cmcube; cSNRs)
(e.g. Plewa 1995; Fabian \& Terlevich 1996). 

In the Antennae we detect 18 ULXs.
Of these,  3 are variable, and 4 are detected with
luminosities above  $10^{40}$\ergs. As discussed in Paper~II,
 sources with $\rm{L_{X}>10^{39}}$~\ergs~  have hard X-ray spectra with
power-law indices $\Gamma \leq2.0$. 
 The two most luminous of them (sources 16 and 42) have
extremely  hard X-ray spectra with photon indices of $\Gamma\sim1.2$.
 For these sources  there is also evidence of gaussian emission lines
(\ion{N}{4} triplet (0.43~keV),  \ion{Mg}{9} triplet
(1.33-1.34~keV)  and \ion{Mg}{13} (1.5~keV)). 
This result is quite surprising given the lack of any Fe-L emission around
1.0~keV and of any thermal continuum in their spectra (Paper~II and \S3.)
and  suggests  either fluorescence or  photoionisation of the ambient
gas   by the intense  emission from the central X-ray source.

Although cSNR models predict hard X-ray spectra, like those observed
in our sources (Plewa 1995; Franco \etal 1993), it is 
unlikely that they account for the majority of our brightest ULXs,
since the latter
do not have clear radio counterparts. 
 Based on
the lightcurve of SN1988Z which is one of the best studied examples of
cSNRs (Aretxaga \etal 2001), we estimate that for an X-ray luminosity
of $\sim10^{40}$~\ergs~ we should detect a radio counterpart with a
flux of $\sim1$~mJy at 6cm, but we do not. Such a source should be detectable in the radio
observations  of NU2000 which have a detection limit of
$\sim50~\mu$Jy at 6~cm. 
Of the lower luminosity sources ($\rm{L_{X}\sim10^{39}-10^{40}}$\ergs) only 5  have 
possible radio counterparts. Four of
them (26, 29, 33,34), which have  steep radio spectra, could still
have some contribution from SNRs. 

X-ray binary (XRB) counterparts are a more likely possibility: the X-ray
spectra of the ULXs in the Antennae can be fit with power-laws in the range of
those observed in XRB spectra (e.g., Van Paradijs 1999) and   their co-added spectrum 
 (\S3.) is well fit with a composite
disk-BB plus power-law model, with comparable strengths in the two components, as seen
in black-hole binaries in the Very High State (VHS; e.g., Esin \etal 1997).
However, at variance with VHS black hole binaries which have cooler
inner disks (e.g., Tanaka \& Shibazaki 1996; however
see Esin \etal 1997 for some exceptions), the inner temperature of the
model accretion disk for these sources
is rather large, $\sim1.2$~keV.
Moreover, the slope of the hard component in the Antennae ULXs  is much flatter
($\Gamma\sim1.2-1.5$) than for most black-hole binaries   in the VHS
($\Gamma\sim2.5$; e.g., Tanaka \& Lewin 1995), suggesting that if ULXs
 are XRBs they probably belong to a different population  than the known 
Galactic black-hole binaries in high/very high state.  

The spectral parameters of the Antennae ULXs are similar to those of Galactic
microquasars (e.g., Miller \etal 2001), suggesting a possible link,
but the luminosities are much larger.
% properties
%of their accretion disks may be very similar, 
From the X-ray spectra of the ULXs we cannot  distinguish between
 X-ray binaries with early type companions (High Mass X-ray Binaries;
HMXBs) and  X-ray binaries with late type companions (Low Mass X-ray
Binaries; LMXBs), since the X-ray
spectra provide information only on 
the nature of the accreting object. However,  association
with young stellar clusters suggests that their majority
is related to HMXBs. 

Both spectral parameters and luminosities of Antennae ULX 
are consistent with the properties of other ULXs,  
including time-variable sources with the characteristic
 high/soft - low/hard transitions observed in black hole binaries 
(e.g., Makishima \etal 2000; LaParola \etal 2001; Kubota
\etal 2001).  If the
ULXs in the Antennae are black hole binaries, with spherical accretion onto the 
black hole, their extreme high X-ray
luminosities suggest black hole masses in the $\gg$10-1000\msun~ range.
As for the other ULXs, the high temperatures we infer
for the inner part of the accretion disk are inconsistent with 
black holes in this mass range, unless rotating black holes
(e.g. Makishima \etal 2000) or slim/thick accretion disks (e.g. Watarai
\etal 2000) are invoked.

An alternative  model suggested for these sources, that of moderate
beaming, would not require such massive black holes (King \etal 2001).
Our comparison of X-ray and optical images of the Antennae (\S3.1)
may support this possibility at least for some sources.
Although 10 out of 14 ULXs in the Antennae observed with the Hubble WFPC2
have possible optical associations (a
stellar cluster down to a limiting magnitude of $\rm{m_{V}=23}$~mag, within a radius of $2\arcsec$), only three can be securely associated with 
a stellar cluster (distance less than $1\arcsec$; Fig. 3d).  
The other 4 of these ULXs may have even more distant optical counterparts.
The $1\arcsec-2\arcsec$ offsets (corresponding to a projected physical distance 
of $\sim300$~pc) between the X-ray and the stellar clusters   
measured for most of these sources are reminiscent of runaway binaries.
Runaway binaries are the result of deviation from spherical symmetry
during SN explosions (see van den Heuvel \etal 2000 for a discussion of these systems).
Van Paradjis \& White (1995)   explain the wide Galactic latitude distributions
of LMXBs with this mechanism. 
\footnote[1]{ However, the galactic black-hole candidate XTE
J1118+480 is located in the halo of the Galaxy. 
It is unclear whether this source was formed in the
Galactic plane and the kicked out, or was  
formed in the halo.} 

Runaway binaries are not likely to be
massive systems, as shown by by the narrower distribution of the
Galactic latitude of LMXBs with black holes compared to those with neutron
star accretors (White \& van Paradijs, 1996), and by SNe models
for massive stars (Arnett, 1996), where only a small fraction of
the stellar mass is
expelled during the explosion, resulting in very small, if any, kicks
to the remnants.
  The possibility that some at least of the ULXs may be runaway binaries is supported by 
the similar offsets
measured in less luminous sources (right panels in Fig.~3d), which are
likely to be normal lower-mass binaries and would be subject to kicks, 
by analogy with the galactic XRB observations. 

The implications of these offsets will be discussed further in Paper IV (Zezas \&
Fabbiano 2002), together with constraints to ULX models derived from the 
luminosity distribution of the Antennae sources.

\subsection{Sources with $\rm{L_{X}\sim10^{38}-10^{39}}$\ergs}

 X-ray spectra of these sources are much less constrained than those of
the more luminous sources, with spectral parameters covering 
a very wide range of values ($1.7<\Gamma<7$, and \nh~
ranging from the galactic value up to
$\sim3\times10^{22}$\cmsq; Paper~II). 
The co-added spectrum of the sources with 
$L_X > 3\times10^{38}$\ergs, which is the Eddington limit for a neutron
star,  is best fit with a power-law ($\Gamma=2.0$)
plus two thermal plasma models (kT=0.76~keV and 0.28~keV). The
parameters of the thermal component are reminiscent of those obtained
for the superwinds in the nearby galaxies M~82 and NGC~253 (Strickland
\etal 1999, 2000; Moran \etal 1999), suggesting that a fraction
of the total emission of these sources is produced by a multicomponent
thermal plasma. The value of the photon index is  steeper than
that of the higher luminosity sources and is at the borderline between
that observed for black-hole binaries in  high state and magnetized neutron star binaries 
in the Galaxy (e.g. van Paradjis, 1999). However, the co-added spectrum
can be fitted equally well with a harder power-law  in the range of
neutron star binaries
($\Gamma=1.5$), a multi-temperature disk-BB (kT=0.24) and a
thermal plasma (kT=0.66).  This  disk-BB temperature 
component is much lower than that of the ULXs
 and is consistent with that observed in VHS Galactic  black-hole binaries
(e.g. Tanaka \& Shibazaki, 1996; Makishima \etal 2000).
The best fit power-law slope, instead, is flatter than that of
Galactic black-hole binaries.

Of the 18 sources in this luminosity range only 10 are securely associated
with a stellar cluster of a limiting magnitude $\rm{m_{V}=23}$~mag.
The V-I colours of these clusters are consistent with ages less than 100~Myr, as
determined from evolutionary synthesis models (Leitherer \etal 1999). 
 Ten of these sources are associated with steep spectrum radio
sources. Thus these
sources are closely connected with the younger stellar population.
 
The lowest luminosity sources (i.e., those with luminosities below 
the Eddington limit for a neutron star binary, 
$\rm{L_{X}\sim3\times10^{38}}$\ergs),
are most probably a mixture
of SNRs and neutron star binaries. This is suggested by the steepness
of their X-ray spectra, which have a mean photon index of 2.5 as measured
from their co-added spectrum. Fits to the same spectrum also require a
thermal component with a temperature of 0.18~keV, most probably
associated with diffuse gas in the neighborhood of the sources. This
component becomes more important in the lowest luminosity sources
since it is not negligible compared to the source's emission. We note that models with
a multi-temperature disk-black body spectrum do not give acceptable
fits, suggesting that black hole binaries in high state are not the dominant
population in this luminosity range. This is expected since black hole binaries in
 high state emit close to their Eddington luminosity which is higher than 
$\sim3\times10^{38}$~\ergs~. Also it is unlikely that these sources are
black-hole binaries in the low state, since in this case their
luminosity would be less than 10\% of their Eddington luminosity
(e.g. Nowak 1995). Indeed the slope of the power-law component is steeper
than the typical slopes of black-hole binaries in low state
($\Gamma\sim1.5-1.7$; e.g. Esin \etal 1997).

  The sources with soft X-ray spectra ($\Gamma>3$) could well be
supernova remnants, and this
explains the excess of softer sources found at lower
luminosities. Indeed,  some
of them have radio counterparts with steep radio spectral indices. 
NU2000 estimate that about 50 SNRs are enough to
produce the observed radio emission for most non thermal radio
sources, consistent with the number of
SNRs required to explain the observed X-ray luminosity of their counterparts.  

 Whitmore \etal (1999)
identified some of the sources detected with  
the HST/WFPC2 as globular clusters. None of these sources are found to
be associated with an X-ray source. This could be due to our 
relatively high detection threshold, since the dominant X-ray 
population in globular clusters is LMXBs which have luminosities close
to or lower than our detection limit of 
$10^{38}$~\ergs.

\subsection{Highly obscured sources}

As shown in Paper~II,  six sources have column densities ranging from $0.35\times10^{22}$\cmsq up to
$3.4\times10^{22}$\cmsq, i.e. $\sim 10 - 100$  times the Galactic
\nh~ along the line of sight.
 These sources (12, 24, 25, 34, 35 and 36),  are
found predominantly in the impact region between the two galaxies and in the neighborhood of
molecular clouds (\S3) (see Fig.~5).  Although the majority of these sources have
low count rates, their intrinsic luminosities can be  high because of their very large \nh.  The most extreme
examples are Sources~35 and 24 with an absorption corrected luminosity
of  $8\times10^{39}$\ergs~  and  $3\times10^{41}$\ergs~ respectively. It is unlikely that both
sources are background QSOs seen
through the dense gaseous disk of the Antennae system, because the \nh~
inferred for the X-ray fits are well below those measured from
observations at 21cm ($4.2\times10^{24}$\cmsq and 
$\sim6.0\times10^{24}$\cmsq for Source~35 and Source~24 respectively;
Gordon  \etal 2001).  This comparison suggests that both sources are situated within the
molecular clouds, and reside in  young stellar clusters, a 
conclusion  supported  by the fact that both sources are coincident with 
strong non-thermal radio sources.  Based on the conservative
 model of Condon \& Yin (1990) NU2000 estimate a SN rate of 5.1 SNe per $10^{3}$~yr 
 for Source~35, which is the strongest non-thermal radio source in the
Antennae. Less conservative estimates give 10 times higher rates.
Observations of other star-forming galaxies (eg M82, Zezas
\etal 2001; NGC3628, Strickland \etal 2001), reinforce the
conclusion that X-ray sources with
luminosities up to $10^{41}$~\ergs~ are not uncommon  in  very active star-forming
regions. Therefore, these sources could be either extreme cases of
ULXs or complexes of luminous X-ray binaries in compact star-forming
region.

%From the luminosity function calculated for the rest of the galaxy we
%can estimate the number of sources which are hidden in the overlap
%region. We find XXX sources with intrinsic luminosities above
%$10^{40}$\ergs in this region. By simply rescaling the LF determined in section 2.5
%we estimate that XXX sources with intrinsic luminosities between
%$10^{38}$ up to $10^{39}$\ergs are situated in this molecular cloud.

\subsection{The northern nucleus: NGC~4038}

 Based on the radio coordinates of the Northern nucleus we identify it
as our source 25. The offsets of the centroid of this source
(0.3-10.0~keV band) from the 4~cm and 6~cm radio sources are $1.1''$ and
$1.87''$, respectively. In this region there are  three more sources
(sources 24, 22, 26), of which the last two are only detected above
2.0~keV.
 This is not surprising, given the large amounts of molecular
gas present in this region (Wilson \etal 2000).  The luminosity of the
nucleus is modest ($\sim10^{39}$\ergs) just slightly higher than the
luminosity of two of the neighboring sources. Its spectrum 
is very soft ($\Gamma=8.26^{+1.74}_{-2.61}$) with a relatively large
absorbing column density ($\rm{N_{H}=1.07^{+0.36}_{-0.42}}\times10^{22}$~\cmsq). The two nearby
 sources have similar X-ray spectra. The steep spectrum
of the nuclear sources suggests thermal emission. The fit with a PO+RS model results in
a temperature of kT$~\sim0.8$~keV (Table~7 of Paper~II), which could be consistent with thermal emission
 by a supernova driven superwind (e.g., Heckman \etal 1980; Strickland \etal 2000). 
This interpretation is supported by the relatively flat radio spectral index of the nucleus
 ($\alpha=-0.58\pm0.20$) and the presence of a number
of young star-clusters in the area. 
A more detailed discussion on this will be presented in
Paper V (Fabbiano \etal in preparation).

\subsection{The southern nucleus: NGC~4039}

 We identify the southern nucleus as our source 29 based on its radio position.
 It has an X-ray
luminosity of $7.7\times10^{39}$\ergs~ and a steep photon index of
$3.3^{+0.5}_{-0.4}$. The inclusion of a thermal component improves
the fit and gives a relatively  absorbed power-law
(\nh $\sim0.14\times10^{22}$\cmsq) with a steep slope ($\Gamma\sim2.14$). The
temperature of the thermal component is 0.77~keV. The existence of the
additional thermal component is consistent  with the fact that an
extended component is associated with this source. These results suggest
that there is a  population of X-ray binaries together with diffuse gas,
possibly from an outflow from an earlier starburst. Indeed, IR
spectroscopic results show that the nucleus of NGC~4039 does not host any recent
star-formation but it has an old stellar population with an age of
$\sim60$~Myr (Mengel \etal 2001).  At this age the majority of the OB type stars have
formed supernovae and most of the intermediate type stars are evolving
off the main sequence. However, it is still too early for the onset of
Roche lobe overflow in late type stars, which initiates the X-ray
emitting phase of LMXBs (eg. Meurs \& Van den Heuvel 1989). Therefore we expect the majority of the
X-ray binaries to be HMXBs (neglecting any underlying LMXBs from the
bulge of NGC~4039). This is a reasonable approximation  since the
stellar populations are now dominated by the most recent
star-formation events. The nucleus of NGC 4039 has a very steep radio spectrum
suggesting that its radio emission is non-thermal, arising from
SNRs, in accordance with an intermediate age starburst. These SNe may be responsible for heating the ISM in the nuclear
region to X-ray emitting temperatures, consistent with the soft thermal component measured in
its spectrum.

\section{Conclusions}

In this paper we have discussed the X-ray properties and the
nature of the luminous X-ray sources discoved with \chandra in the Antennae (Paper~I),
and analyzed in detail in Paper~II, by studying the average properties of
the X-ray spectra in three luminosity ranges, and by investigating 
the multi-wavelength environment of the X-ray sources.
Our results are summarized below:

\medskip
\noindent
1)
As reported in Paper~II,
49 sources were detected with \chandra in the Antennae, down to
a limiting luminosity of $10^{38} \rm ergs~s^{-1}$ ($H_o = 50$);
18 of these sources are ULXs with $L_X > 10^{39} \rm ergs~s^{-1}$.
At least 4 of these sources are variable either with the \chandra
observations, or in comparison with the previous {\it ROSAT HRI}
observation (Fabbiano \etal 1997). The most luminous ULXs have
very hard spectra, that can be fitted with power-laws of $\Gamma
\sim 1.2$, and there is evidence of a luminosity-spectral trend,
with sorces becoming softer at the lower luminosities.

\medskip
\noindent
2)
We co-added the X-ray spectra of sources in 3 luminosity ranges:
$L_X > 10^{39} \rm ergs~s^{-1}$ (ULXs), $3 \times 10^{38} < L_X < 10^{39} \rm ergs~s^{-1}$,
and $L_X < 3 \times 10^{38}\rm ergs~s^{-1}$, and fitted the co-added
spectra with a number of emission models. Single-component
models are inadequate to fit the data in all cases.

The spectrum of sources with  $L_X < 3 \times 10^{38}\rm ergs~s^{-1}$ is well fitted
with a composite Power-law ($\Gamma \sim 4$) and thermal Raymond-Smith 
(kT$\sim$ 0.4~keV) model, with the
thermal component accounting for 34\% of the emission at energies below 2.5~keV. 
A good fraction of this component is likely to arise from hot ISM
in the beam.

Sources with $L_X >3 \times 10^{38} \rm ergs~s^{-1}$ require three-component
models, either a power-law with two Raymond-Smith components, or a
powerlaw + Raymond-Smith + disk-BB. The latter model (multi-color disk
black body) has been used to approximate the emission of an 
accretion disk in X-ray binaries (e.g. see Makishima \etal 2000).

In sources with $3 \times 10^{38} < L_X < 10^{39} \rm ergs~s^{-1}$,
the Raymond-Smith component accounts for a sizeable amount
of the soft emission. Again, it is possible that this may be due
at least in part to hot ISM in the beam. If we adopt the power-law
plus two thermal components model, we obtain power-law $\Gamma \sim 2$,
consistent with either black-hole binaries or magnetized 
neutron star binaries. The temperatures of the thermal components
($\sim$0.8 and $\sim$0.3~keV), are in the range of those measured in
the hot ISM and superwinds of nearby starburst galaxies (e.g. Strickland 
et al 1999). If we adopt the powerlaw + Raymond-Smith + disk-BB model,
the power-law is flatter, $\Gamma \sim 1.5$, the plasma temperature is
still in the range expected for a hot ISM, and the inner accretion disk
temperature is in the range of those seen in Very High State Galactic
black-hole binaries (kT$\sim$0.2~keV; e.g. Tanaka \& Shibazaki 1996).

In the ULX spectrum, the Raymond-Smith contribution to the emission is small.
Adopting the powerlaw + Raymond-Smith + disk-BB model, the best-fit
parameters are consistent with those of other ULXs (e.g. Makishima \etal 2000,
LaParola et al 2001; Kubota \etal 2001). In particular the inner accretion disk
temperature is large (1.13~keV), requiring a rotating massive black hole, if
the accreting black hole binary picture of Makishima \etal (2000) applies.

\medskip
\noindent
3)
 Twenty three sources  are found to be coincident with one or more young stellar clusters
(ages $<30$~Myr) and one with an intermediate age cluster.  This
suggests that they are most probably associated with a young stellar
population.   This is also supported by the IR spectra of the four
clusters with available IR spectroscopy.  
Comparison with {\it HST} data (Witmore et al 1995, 1997) shows that the majority
of sources have an associate optical cluster within a 2" radius. However, there
are random orientation offsets between X-ray source and optical cluster
position in a number of cases, including ULXs. This suggests that, if the
X-ray source originated in the star cluster, it was ejected at a certain 
point in its evolution, i.e. it is a runaway binary. If ULXs are runaway binaries,
it is unlikely that they may be very massive. This results therefore
challenges the association of ULXs with 10-1000 \msun~ black holes, while is in
keeping with the moderate beaming model of King \etal (2001). 

\medskip
\noindent
4)
Comparison with the Neff \& Ulvestad (2000) radio continuum survey of the Antennae,
shows that twenty two X-ray sources have one
or more radio counterparts. The majority of these have steep radio
indices suggesting that SNRs are responsible at least for part of
their soft X-ray emission.  
While fainter sources may be associated with SNRs, the ULXs are not 
likely to be, reinforcing the binary model for these sources.
in particular, the lack of radio counterparts for the 3
most luminous X-ray sources ($\rm{L_{X}>10^{40}}$~\ergs) strongly
indicates that they are not associated with young compact SNRs.

\medskip
\noindent
5) 
We find 6 sources with obscuring
column densities more than 10 times higher than the Galactic along the
line of sight. These sources are spatially coincident with dense
molecular clouds suggesting that they are regions of intense
star-formation embedded in these clouds. 

\medskip
\noindent
6)
Finally, both nuclei were detected. The northern nucleus has $L_X \sim 10^{39}
\rm ergs~s^{-1}$, the southern nucleus
is associated with extended emission and has $L_X \sim 8 \times 10^{39}
\rm ergs~s^{-1}$. The northern nucleus has a very soft spectrum, suggesting
thermal emission, possibly related to hot winds escaping the star
forming region. The southern nucleus has a composite spectrum, consisting
of a thermal component plus a power-law with $\Gamma \sim 2$. This spectrum
suggests both hot ISM and a contribution with X-ray binaries, which
would be expected to be found in this post-starburst nucleus
(age of $\sim 60$Mys; Mengel \etal 2001).

\medskip
These results clearly need to be followed up by more detailed 
deep observations of the Antennae with \chandra, so that individual
sources can be studies with adequate statistics, instead of relying 
only on average properties. A `large' \chandra observing program 
to pursue these objectives was approved in AO-3, and the data are
beginning to be acquired.

\acknowledgments

 We thank the CXC DS and SDS teams for their efforts in reducing the data and 
developing the software used for the reduction (SDP) and analysis
(CIAO). We thank Martin Ward, Jeff McClintock, Andrea Prestwich and Phil Kaaret for
useful discussions on these results. We also thank J. Hibbard for providing the HI VLA
data for the Antennae. The HST data presented in this paper were obtained from the
Multimission Archive at the  Space Telescope Science Institute (MAST).
This work was supported by NASA contract NAS~8--39073 (CXC)
and NAS8-38248 (HRC).

\clearpage

%%%%%%% TABLE-9 %%%%%%%%%%%%%%%%%%%%%

\makeatletter
\def\jnl@aj{AJ}
\ifx\revtex@jnl\jnl@aj\let\tablebreak=\nl\fi
\makeatother
\ptlandscape
\begin{deluxetable}{lccccccccccc}
\tabletypesize{\scriptsize}
\rotate
\tablecolumns{11}
\tablewidth{0pt}
\tablenum{1A}
\tablecaption{ Spectral Fits of Coadded spectra }
\tablehead{ \colhead{$\rm{L_{X}}$ range}& \multicolumn{2}{c}{PO} &
\multicolumn{2}{c}{PO+RS} &   \multicolumn{2}{c}{PO + disk bb } &
\multicolumn{2}{c}{PO + RS + disk bb} & \multicolumn{2}{c}{PO + RS +
RS}  \\
\colhead{ }	&\colhead{ $\Gamma$} &\colhead{ $\rm{N_{H}}^{1}$} &
\colhead{ $\Gamma$} &\colhead{kT$^{2}$ } &  \colhead{ $\Gamma$} &
\colhead{ $\rm{kT_{in}}^{2}$} & \colhead{ $\Gamma$} &
\colhead{ kT$^{2}$}  & \colhead{ $\Gamma$} &
\colhead{ kT$^{2}$} \\
\colhead{ }	&\colhead{ } &\colhead{ \x2}  &\colhead{$\rm{N_{H}}^{1}$ }&\colhead{ \x2} 
 &  \colhead{$\rm{N_{H}}^{1}$ } & \colhead{ \x2} &
\colhead{$\rm{kT_{in}}^{2}$}  &  
\colhead{$\rm{N_{H}}^{1}$ } &\colhead{ kT$^{2}$} &\colhead{$\rm{N_{H}}^{1}$ }\\
\colhead{ }	&\colhead{ } &\colhead{ } & \colhead{ } &\colhead{ } 
 &  \colhead{ } & \colhead{ } &  \colhead{}  &  \colhead{\x2 } &\colhead{ } &\colhead{ \x2}\\
\colhead{ (1) }	&\colhead{ (2)}
&\colhead{(3)} &\colhead{(4)} &\colhead{(5)} & \colhead{ (6) }	&\colhead{ (7)}
&\colhead{(8)} &\colhead{(9)} &\colhead{(10)} &\colhead{(11)}}
\startdata
$<3\times10^{38}$	& $	3.36	_{-	0.28	}^{+	0.66	} $ &	$	0.28	_{-	0.05	}^{+	0.1	} $ &\bf{ $	2.5	_{-	0.31	}^{+	0.57	} $ }& \bf$	0.36	_{-	0.09	}^{+	0.12	} $ & $	3.76	_{-	0.46	}^{+	0.56	} $ & $		0.068	_{-	0.002	}^{+	0.012	} $ & $						 $ & $		$ & $	$ & $		_{		}^{		} $ \\[0.1cm]
	& 						&		55.9 (22)					& $	0.18	_{-	0.07	}^{+	0.11	} $ & \bf	16.6 (19)					 & $	1.105	_{-	0.06	}^{+	0.14	} $ & 		31.9 (20)					 & $ $ & $ $ & $		 $ & 	$ $ 					\\[0.1cm]

$3\times10^{38} - 10^{39}$	& $	3.37	_{-	0.26	}^{+	0.32	} $ &	$	0.23	_{-	0.034	}^{+	0.052	} $ & $	2.25	_{-	0.18	}^{+	0.22	} $ & $	0.66	_{-	0.05	}^{+	0.04	} $ & $	1.8	_{-	0.26	}^{+	0.28	} $ & $		0.15	_{-	0.02	}^{+	0.03	} $ & $	1.54	_{-	0.26	}^{+	0.41	} $ & $	0.66	_{-	0.05	}^{+	0.06	} $ &\bf{ $	2.02	_{-	0.18	}^{+	0.24	} $} &\bf {$	0.76	_{-	0.07	}^{+	0.07	} $} \\[0.1cm]
	& 						&		256.7 (85)					& $	0.08	_{-	0.02	}^{+	0.02	} $ & 	124.1 (83)					 & $	0.32	_{-	0.06	}^{+	0.11	} $ & 		160.3 (83)			&$			0.24	_{-	0.06	}^{+	0.05	} $ & $	0.09	_{-	0.03	}^{+	0.06	} $ &\bf {$	0.28	_{-	0.03	}^{+	0.06	} $ }&\bf {$	0.072	_{-	0.02	}^{+	0.03	} $ }\\[0.1cm]
	& 						&							& 						  & 						 & 						&							&					&	107.2 (81)					&						&\bf	99.7 (81)					\\[0.1cm]
$>10^{39}$	& $	1.57	_{-	0.04	}^{+	0.05	} $ &	$	0.09	_{-	0.01	}^{+	0.04	} $ & $	1.47	_{-	0.04	}^{+	0.04	} $ & $	0.82	_{-	0.06	}^{+	0.15	} $ & $	1.34	_{-	0.19	}^{+	0.18	} $ & $		0.60	_{-	0.2	}^{+	0.27	} $ &\bf {$	1.23	_{-	0.27	}^{+	0.23	} $ }&\bf{ $	0.81	_{-	0.08	}^{+	0.05	} $ }& $	1.55	_{-	0.06	}^{+	0.06	} $ & $	0.77	_{-	0.09	}^{+	0.07	} $ \\[0.1cm]
	& 						&		281.4 (244)					& $	0.078	_{-	0.01	}^{+	0.008	} $ & 	247.6 (242)					 & $	0.066	_{-	0.018	}^{+	0.022	} $ & 		276.0 (242)			&\bf$			1.13	_{-	0.24	}^{+	0.47	} $ &\bf{ $	0.042	_{-	0.017	}^{+	0.019	} $ }& $	0.054	_{-	0.012	}^{+	0.047	} $ & $	0.12	_{-	0.03	}^{+	0.02	} $ \\[0.1cm]
	& 						&							& 						  & 						 & 						&							&					&\bf	229.7 (240)					&		&						232.5 (240)					\\[0.1cm]
\enddata
\tablenotetext{1}{ Absorbing column density in units of $10^{22}~\rm{cm^{-2}}$. }
\tablenotetext{2}{ Temperature in units of keV. }
\end{deluxetable}

\makeatletter
\def\jnl@aj{AJ}
\ifx\revtex@jnl\jnl@aj\let\tablebreak=\nl\fi
\makeatother
\ptlandscape
\begin{deluxetable}{lccccccccccc}
\tabletypesize{\scriptsize}
%\rotate
\tablecaption{ Intensity of the Best Fit Components in the  Coadded Spectra}
\tablecolumns{17}
\tablewidth{0pt}
\tablenum{1B}
\tablehead{ \colhead{$\rm{L_{X}}$ range}&
\colhead{Model}&\multicolumn{2}{c}{Intensity$^{1}$} \\
\colhead{ }& \colhead{Component} &  \colhead{(0.1-2.5) keV} &  \colhead{(2.5-10.0) keV}\\
\colhead{ (1) }	&\colhead{ (2)}
&\colhead{(3)} &\colhead{(4)}}
\startdata
$<3\times10^{38}$	& PO & 0.66 & 0.999 &	\\
			& RS & 0.34 & 0.001 &      \\
\hline
$3\times10^{38} - 10^{39}$  & PO           & 0.63 & 0.98  \\
 			    & RS (high kT) & 0.27 & 0.01 \\
		            & RS (low kT)  & 0.1 & 0.01 \\
\cline{2-4}
			    & PO      & 0.47 & 0.99   \\       
			    & disk-BB & 0.07 & 0.005  \\       
			    & Ray     & 0.46 & 0.005    \\
\hline
$>10^{39}$	& PO      & 0.55 & 0.823 & 	\\
		& disk-BB & 0.38 & 0.175 &       \\
		& Ray     & 0.07 & 0.002 &      \\
\cline{2-4}
                & PO           & 0.91 & 0.999	  \\
 		& RS (high kT) & 0.06 & 0.001 \\
		 & RS (low kT)  & 0.03 & 0.0 \\
\enddata
\tablenotetext{1}{ Relative contribution of each component in the
total emission in each band.}
\end{deluxetable}

%%%%%%% TABLE-10 %%%%%%%%%%%%%%%%%%%%%

\ptlandscape
\begin{deluxetable}{lcccccccccccccccc}
\tabletypesize{\scriptsize}
\tablecolumns{10}
\tablewidth{0pt}
\tablenum{2a}
\tablecaption{Sources with Bright Optical Counterparts}
\tablehead{ 
\colhead{X-ray} &  \colhead{$\Gamma$} &\multicolumn{2}{c}{Log($\rm{L_{x}}$)}& \colhead{Optical}  &  \colhead{Dist.} &\colhead{Mv} &  \colhead{U-B}  &
\colhead{B-V}&  \colhead{V-I}&  \colhead{Notes} \\
\colhead{Source No } & \colhead{ } &  \colhead{soft} & \colhead{hard} &\colhead{Source No }& \colhead{(arcsec)}\\ 
\colhead{ (1) }	&\colhead{ (2)}
&\colhead{(3)} &\colhead{(4)} &\colhead{(5)} & \colhead{ (6)}
&\colhead{(7)} &\colhead{(8)} &\colhead{(9)} & \colhead{ (10)}&
\colhead{ (11)}  }
\startdata
5  & $ 7.22_{-1.71}^{+2.39} $& 39.02 & 37.46 & 40 & 2.3 & -11.46 & -0.7 & 0.06 & 0.17 & young \\
5  & $                      $&       &       & 46 & 2.6 & -11.37 & -0.79 & -0.19 & -0.15& young \\
\hline
6  & $ 4.95_{-1.51}^{+3.64} $& 38.78 & 37.45 & 12 & 0.7 & -12.32 & -0.71 & 0.0  & 0.12 & young \\
6  & $                      $&       &       & 47 & 0.8 & -11.37 & -0.68 & 0.05 & 0.12 & young \\
6  & $                      $&       &       & 50 & 0.6 & -11.33 & -0.8 & -0.05 & 0.22 & young \\
\hline
7  & $                      $&       &       & 22 & 2.4 & -11.78 & -0.69 & 0.04 & 0.27 & young \\
\hline
10  & $ >5.94               $& 38.75 & 37.06 & 1  & 2.0 & -13.92 & -0.61 & 0.02 & 0.49 & young \\
10  & $                     $&       &       & 44 & 2.4 & -11.38 & -0.31 & 0.58 & 1.15 & young \\
\hline
11  & $ 1.54_{-0.18}^{+0.19}$& 39.47 & 39.81 & 49 & 1.5 & -11.34 & -0.56 & 0.28 & 0.32 & young \\
\hline
13  & $ 4.19_{-1.22}^{+1.84}$& 37.24 & 38.66 & 7  & 2.3 & -9.83  & 0.1 & 0.27 & 0.51 & interm \\
\hline
15  & $                     $&\multicolumn{2}{c}{38.40$^1$} & 25 & 0.3 & -11.75 & -0.7 & 0.26 & 0.42 & young \\
15  & $                     $&       &       & 23 & 3.3 & -11.78 & -0.71 & 0.09 & 0.19 & young \\
\hline
22   & $                    $& 38.38 & 38.18 & 3  & 2.1 & -13.68 & -0.72 & 0.06  & 0.06 & young \\
\hline
24  & $ 5.73_{-1.35}^{+2.16}$& 38.88 & 37.77 & 38 & 2.0 & -11.5  & -0.07 & 0.82 & 1.0 & young \\
24  & $                     $&       &       & 34 & 2.2 & -11.61 & -0.24 & 0.64 & 0.84 & young \\
\hline
25  & $ 8.26_{-2.61}^{+1.74}$& 38.63 & 37.35 & 3  & 1.4 & -13.68 & -0.72 & 0.06 & 0.06 & young \\
25  & $                     $&       &       & 4  & 1.9 & -12.9  & -0.67 & 0.05 & 0.07 & young \\
\hline
30  & $                     $&       &       & 13 & 2.4 & -9.41  & 0.99 & 1.42 & 2.04 & interm \\
\hline
34  & $ 6.48_{-3.52}^{+0.62}$& 39.06 & 38.15 & 39 & 0.59 & -11.48& -0.68 & 0.15 & 0.24 & young \\
34  & $                     $&       &       & 30 & 0.6 & -11.67 & -0.69 & 0.14 & 0.14 & young \\
34  & $                     $&       &       & 33 & 0.9 & -11.63 & -0.67 & 0.17 & 0.09 & young \\
34  & $                     $&       &       & 26 & 1.0 & -11.75 & -0.55 & 0.31 & 0.73 & young \\
34  & $                     $&       &       & 14 & 1.3 & -12.2  &  -0.79 & 0.37 & 0.28 & young \\
34  & $                     $&       &       & 9  & 1.3 & -12.38 & -0.73 & 0.39 & 0.34 & young \\
34  & $                     $&       &       & 20 & 1.5 & -11.95 & -0.65 & 0.53 & 0.46 & young \\
34  & $                     $&       &       & 11 & 1.6 & -12.34 & -0.28 & 0.84 & 0.67 & young \\
\hline
36  & $  >3.71              $&\multicolumn{2}{c}{38.0$^1$}& 18 & 1.4 & -12.03 & -0.67 & 0.25 & 0.31 & young \\
\enddata
\tablenotetext{1}{ The luminosity for these sources is in the 0.1-10.0
keV band (from Table~1).}
%$^{2}$ Log of the luminosity (corrected for galactic absorption) in units of erg/s.\\
%$^{3}$ The 4-6cm radio spectral index. Indices measured for extended sources including more than one point sources are marked with an asterisk.\\
\end{deluxetable}

%%%%%%% TABLE-10B %%%%%%%%%%%%%%%%%%%%%

\ptlandscape
\begin{deluxetable}{lcccccccccc}
\tabletypesize{\scriptsize}
\tablecolumns{10}
\tablewidth{0pt}
\tablenum{2b}
\tablecaption{Optical Properties of the Sources with Faint Optical Counterparts}
\tablehead{ 
\colhead{X-ray} & \colhead{Optical} & \colhead{Dist.} &\colhead{Vmag} &  \colhead{V-I} & \colhead{} & \colhead{X-ray} & \colhead{Optical} & \colhead{Dist.} &\colhead{Vmag} &   \colhead{V-I}\\
\colhead{Source } &\colhead{Source }&\colhead{(arcsec) } & \colhead{} &\colhead{} &\colhead{} & \colhead{Source } &\colhead{Source }&\colhead{(arcsec) } & \colhead{}\\
\colhead{ (1) }	&\colhead{ (2)}
&\colhead{(3)} &\colhead{(4)} &\colhead{(5)} & & \colhead{ (6) } &\colhead{ (7)}
&\colhead{(8)} &\colhead{(9)} &\colhead{(10)}  }
\startdata
5   &  347 &  0.6  &  21.12  &  -0.04 & &  25 &  414  &  1.1  &  21.84  &  1.03   \\
5   &  345 &  0.9  &  20.34  &  0.44  & &  25 &  442  &  1.4  &  17.46  &  0      \\
5   &  352 &  1.6  &  21.88  &        & &  25 &  411  &  1.4  &  21.57  &  0.29   \\
5   &  377 &  2.0  &  22.31  &  1.6   & &  25 &  412  &  1.8  &  21.5   &  0.25   \\
5   &  329 &  2.0  &  21.59  &        & &  25 &  403  &  1.9  &  21.93  &  0.74   \\ \cline{1-5}
6   &  479 &  0.2  &  20.05  &  -0.16 & &  25 &  450  &  1.9  &  18.25  &  0.05   \\ \cline{7-11}  
6   &  483 &  0.7  &  21.39  &        & &  29 &  40   &  1.3  &  22.41  &         \\
6   &  485 &  0.7  &  18.72  &  0.21  & &  29 &  49bc &  2.0  &  17.71  &  1.57   \\ \cline{7-11}  
6   &  481 &  0.8  &  19.47  &  0.06  & &  31 &  494  &  1.4  &  21.69  &  0.08   \\
6   &  487 &  0.9  &  21.36  &  0.54  & &  31 &  498  &  1.6  &  19.85  &  0.28   \\ \cline{7-11}  
6   &  491 &  1.9  &  20.56  &  0.46  & &  33 &  65   &  0.7  &  21.94  &  0.33   \\ \cline{1-5}
7   &  296 &  1.3  &  22.43  &  0.86  & &  33 &  55   &  1.8  &  22.73  &         \\ \cline{7-11}  
7   &  297 &  1.4  &  19.54  &  0.29  & &  34 &  87   &  0.6  &  19.52  &  0.65   \\  
7   &  292 &  1.8  &  20.17  &  0.67  & &  34 &  86   &  0.7  &  20.17  &  0.55   \\  
7   &  284 &  1.9  &  19.96  &  0.36  & &  34 &  90   &  1.1  &  18.59  &  0.59   \\ \cline{1-5}
10  &  573 &  0.9  &  22.1   &  1.02  & &  34 &  88   &  1.1  &  19.36  &  0.87   \\  
10  &  581 &  1.0  &  22.23  &        & &  34 &  89   &  1.3  &  18.52  &  0.92   \\  
10  &  571 &  1.5  &  20.53  &  0.74  & &  34 &  75   &  1.6  &  20.8   &  0.76   \\ 
10  &  580 &  1.8  &  23.2   &        & &  34 &  77   &  1.6  &  21.8   &         \\ \cline{7-11}  
10  &  562 &  2.0  &  21.83  &  0.23  & &  35 &  109  &  0.4  &  22.77  &         \\ \cline{1-5}
11  &  244 &  0.7  &  21.77  &  0.53  & &  35 &  105  &  1.5  &  22.47  &         \\ \cline{7-11}  
11  &  235 &  1.0  &  20.94  &  0.29  & &  36 &  359  &  1.1  &  22.74  &  1.89   \\
11  &  253 &  1.6  &  19.73  &  0.52  & &  36 &  384  &  1.2  &  21.29  &         \\ \cline{1-5}
15  &  233 &  0.7  &  21.79  &        & &  36 &  389  &  1.4  &  19.27  &  0.39   \\
15  &  229 &  0.7  &  20.18  &  0.83  & &  36 &  392  &  1.8  &  20.29  &  1.02   \\
15  &  236 &  0.8  &  19.07  &  0.24  & &  36 &  382  &  2.0  &  21.99  &  0.65   \\ \cline{7-11}  
15  &  218 &  1.2  &  21.57  &  0.47  & &  37 &  38   &  1.7  &  20.59  &  1.48   \\ \cline{7-11}  
15  &  243 &  1.3  &  20.89  &  0.35  & &  39 &  124a &  1.3  &  22.94  &  19.68  \\ \cline{7-11}  
15  &  230 &  1.3  &  21.85  &        & &  40 &  114  &  0.3  &  22.29  &  0.82   \\
15  &  217 &  1.5  &  22.44  &  1.2   & &  40 &  116  &  1.4  &  22.3   &  0.51   \\ \cline{7-11}  
15  &  226 &  1.9  &  20.17  &  0.29  & &  41 &  161  &  1.1  &  21.52  &  1.38   \\ \cline{7-11} \cline{1-5}
16  &  698 &  1.7  &  21.89  &  0.35  & &  42 &  362  &  0.3  &  20.44  &  0.44   \\
16  &  695 &  1.7  &  22.1   &  0.68  & &  42 &  370  &  0.7  &  21.69  &  0.83   \\ \cline{1-5}
17  &  556 &  1.8  &  22.51  &  1.11  & &  42 &  374  &  0.9  &  23.05  &         \\ \cline{1-5}
19  &  36  &  2.0  &  21.59  &  0.87  & &  42 &  348  &  1.4  &  21.65  &  0.79   \\ \cline{1-5}
20  &  717 &  1.4  &  21.84  &  0.27  & &  42 &  368  &  1.6  &  22.03  &         \\ \cline{7-11}
20  &  720 &  1.7  &  22.17  &  0.46  & &  \\ \cline{1-5}
22  &  412 &  0.9  &  21.5   &  0.25  & &  \\ \cline{1-5}
24  &  493 &  0.4  &  19.99  &  0.39  & &  \\
24  &  489 &  0.5  &  20.3   &  1.19  & &  \\
24  &  492 &  1.0  &  20.79  &  1.61  & &  \\
24  &  480 &  1.3  &  20.54  &        & &  \\
24  &  500 &  1.5  &  21.3   &        & &  \\
24  &  501 &  1.8  &  20.21  &  1.72  & &  \\ \cline{1-5}
\enddata               
\end{deluxetable}

%%%%%%% TABLE-11 %%%%%%%%%%%%%%%%%%%%%

\makeatletter
\def\jnl@aj{AJ}
\ifx\revtex@jnl\jnl@aj\let\tablebreak=\nl\fi
\makeatother
\ptlandscape
\begin{deluxetable}{lccccccccccccccc}
\tabletypesize{\scriptsize}
\tablecolumns{11}
\tablewidth{0pt}
\tablenum{3}
\tablecaption{ Radio Properties of the Sources}
\tablehead{
\multicolumn{2}{c}{Source} &   \multicolumn{2}{c}{6 cm} & 
\multicolumn{2}{c}{4 cm} & \multicolumn{2}{c}{X-ray Lumin.} &
\colhead{$\Gamma$} & \colhead{Log($\rm{L_{radio}}$)} &  \colhead{Radio index} \\ 
\colhead{X-ray}  & \colhead{Rad.} & \colhead{Dist.} & \colhead{ Flux } & \colhead{Dist.}
& \colhead{ Flux } & \colhead{Soft} & \colhead{Hard} &
\colhead{ } & \colhead{ }  & \colhead{ }\\
\colhead{ }  & \colhead{ }  & \colhead{(arcsec)} & \colhead{($\mu$Jy)} & \colhead{(arcsec)}
& \colhead{($\mu$Jy) } & \colhead{ } &\colhead{ }
&\colhead{ } &\colhead{($\rm{erg/s/Hz}$)} &\colhead{ } \\
\colhead{(1)}	&\colhead{(2)}
&\colhead{(3)} &\colhead{(4)} &\colhead{(5)} & \colhead{ (6) }	&\colhead{ (7)}
&\colhead{(8)} &\colhead{(9)} &\colhead{(10)}&\colhead{(11)} }
\startdata
5	&	11-2	&	2.1	& $	90	\pm	19	$ &	2.6	& $	64	\pm	10	$ &	39.02	&	37.46	&	7.22	&	25.96	& $				$ \\	
5	&	11-3	&	2.8	& $	355	\pm	29	$ &	2.5	& $	147	\pm	10	$ &		&		&		&	26.55	& $				$ \\	
5	&	11-Total	&		& $	445	\pm	35	$ &		& $	290	\pm	17	$ &		&		&		&	26.65	& $	-0.77	\pm	0.22	$ \\	
\hline
6	&	12-1	&	2.3	& $	55	\pm	11	$ &	1.5	& $	70	\pm	10	$ &	38.78	&	37.45	&	4.95	&	25.74	& $				$ \\	
6	&	12-2	&	1.0	& $	80	\pm	11	$ &	0.5	& $	41	\pm	10	$ &		&		&		&	25.91	& $				$ \\	\cline{4-4} \cline{6-6}
6	&	12-Total	&		& $	135	\pm	16	$ &		& $	111	\pm	14	$ &		&		&		&	26.13	& $	-0.35	\pm	0.34	$ \\	
\hline
7	&	10-1	&	2.0	& $	48	\pm	11	$ &	2.6	& $	56	\pm	10	$ &\multicolumn{2}{c}{38.5$^1$}		&		&	25.68	& $				$ \\	
7	&	10-2	&	2.7	& $	217	\pm	30	$ &	2.5	& $	75	\pm	10	$ &		&		&		&	26.34	& $				$ \\	\cline{4-4} \cline{6-6}
7	&	10-Total	&		& $	265	\pm	32	$ &		& $	131	\pm	14	$ &		&		&		&	26.43	& $	-1.27	\pm	0.32	$ \\	
\hline
10	&	13-4	&	1.9	& $	47	\pm	11	$ &		& $				$ &	38.75	&	37.06	&	$>7.6$	&	25.67	& $	<-0.81	\pm	0.74	$ \\	
10	&	13-5	&	1.8	& $	62	\pm	11	$ &		& $				$ &		&		&		&	25.79	& $	<-1.30	\pm	0.68	$ \\	
\hline
13	&	8-1	&	2.5	& $	48	\pm	11	$ &		& $				$ &	37.24	&	38.66	&	4.19	&	25.68	& $	<-0.84	\pm	0.73	$ \\
\hline											  						          	
15	&	8-2	&	0.9	& $	113	\pm	11	$ &	0.99	& $	90	\pm	10	$ &\multicolumn{2}{c}{38.40$^1$}	&	4.41	&	26.06	& $	-0.41	\pm	0.29	$ \\
15	&	8-3	&	2.6	& $	72	\pm	11	$ &		& $				$ &		&		&		&	25.86	& $	<-1.57	\pm	0.66	$ \\
\hline
17	&	7-13	&	2.1	& $	48	\pm	11	$ &		& $				$ &\multicolumn{2}{c}{38.12$^1$}	&	&	25.68	& $	<-0.84	\pm	0.73	$ \\
\hline
22	&	7-2	&	2.3	& $	145	\pm	11	$ &	2.1	& $	174	\pm	10	$ &	38.38	&	38.18	&	2.44	&	26.15	& $				$ \\
22	&	7-3	&	1.0	& $	96	\pm	11	$ &		& $				$ &		&		&		&	25.98	& $				$ \\	\cline{4-4} \cline{6-6}
22	&	7-Total	&		& $	241	\pm	16	$ &		& $	174	\pm	10	$ &		&		&		&	26.38	& $	-0.58	\pm	0.2	$ \\	
\hline
23	&	1A-2	&	2.8	& $	44	\pm	11	$ &		& $				$ &\multicolumn{2}{c}{38.25$^1$}	&	&	25.64	& $	-0.69	\pm	0.75	$ \\
\hline													
24	&	7-10	&	1.0	& $	1354	\pm	23	$ &	0.9	& $	659	\pm	10	$ &	38.88	&	37.77	&	5.73	&	27.13	& $				$ \\	
24	&	7-11	&		& $				$ &	1.5	& $	275	\pm	10	$ &		&		&		&		& $				$ \\	
24	&	7-12	&		& $				$ &	2.8	& $	104	\pm	10	$ &		&		&		&		& $				$ \\	\cline{4-4} \cline{6-6}
24	&	7-Total	&		& $	1354	\pm	23	$ &		& $	1038	\pm	18	$ &		&		&		&	27.13	& $	-0.48	\pm	0.13	$ \\	
\hline
25	&	7-2	&	1.1	& $	145	\pm	11	$ &	1.87	& $	174	\pm	10	$ &	38.63	&	37.35	&	8.26	&	26.16	& $				$ \\	
25	&	7-3	&	1.2	& $	96	\pm	11	$ &		& $				$ &		&		&		&	25.98	& $				$ \\	\cline{4-4} \cline{6-6}
25	&	7-Total	&		& $	241	\pm	16	$ &		& $	174	\pm	10	$ &		&		&		&	26.38	& $	-0.58	\pm	0.2	$ \\	
25	&	7-7	&	2.9	& $	204	\pm	11	$ &	2.8	& $	153	\pm	10	$ &		&		&		&	26.31	& $	-0.52	\pm	0.2	$ \\	
\hline
26	&	7-7	&	1.7	& $	204	\pm	11	$ &	1.5	& $	153	\pm	10	$ &\multicolumn{2}{c}{39.02$^1$}	&	1.34	&	26.31	& $	-0.52	\pm	0.2	$ \\	
26	&	7-8	&	2.0	& $	109	\pm	11	$ &	2.2	& $	77	\pm	10	$ &		&		&		&	26.04	& $	-0.62	\pm	0.32	$ \\	
\hline
29	&	1-2	&	0.7	& $	513	\pm	11	$ &	0.8	& $	295	\pm	10	$ &	39.53	&	39.19	&	3.33	&	26.71	& $				$ \\	
29	&	1-3	&	2.1	& $	226	\pm	11	$ &	2.0	& $	224	\pm	10	$ &		&		&		&	26.36	& $				$ \\	\cline{4-4} \cline{6-6}
29	&	1-Total	&		& $	739	\pm	16	$ &		& $	519	\pm	14	$ &		&		&		&	26.87	& $	-0.63	\pm	0.14	$ \\	
29	&	1-5	&	3.7	& $	365	\pm	21	$ &	3.9	& $	327	\pm	25	$ &		&		&		&	26.56	& $	-0.2	\pm	0.21	$ \\	
\hline
33	&	2-2	&	1.4	& $	241	\pm	31	$ &	1.8	& $	180	\pm	28	$ &	38.78	&	38.54	&	3.33	&	26.38	& $	-0.52	\pm	0.38	$ \\	
\hline
34	&	2-2	&	2.2	& $	21	\pm	31	$ &	1.9	& $	180	\pm	28	$ &	39.06	&	38.15	&	6.48	&	25.39	& $	-0.52	\pm	0.38	$ \\	
34	&	2-6	&	0.4	& $	2257	\pm	20	$ &	0.4	& $	1957	\pm	21	$ &		&		&		&	27.36	& $				$ \\	
34	&	2-7	&	2.2	& $	59	\pm	11	$ &		& $				$ &		&		&		&	25.77	& $				$ \\	\cline{4-4} \cline{6-6}
34	&	2-Total	&		& $	2316	\pm	20	$ &		& $	1957	\pm	21	$ &		&		&		&	27.37	& $	-0.3	\pm	0.13	$ \\	
34	&	2-8	&	4.7	& $	49	\pm	11	$ &	3.5	& $	41	\pm	10	$ &		&		&		&	25.69	& $	-0.32	\pm	0.61	$ \\	
\hline
35	&	3-5	&	1.3	& $	1121	\pm	19	$ &	1.3	& $	930	\pm	22	$ &	38.33	&	39.63	&	1.24	&	27.05	& $				$ \\	
35	&	3-6	&	1.3	& $	2291	\pm	49	$ &	1.1	& $	1274	\pm	46	$ &		&		&		&	27.36	& $				$ \\	\cline{4-4} \cline{6-6}
35	&	3-Total	&		& $	3412	\pm	53	$ &		& $	2204	\pm	51	$ &		&		&		&	27.54	& $	-0.78	\pm	0.14	$ \\	
\hline
36	&	5-4	&	2.0	& $	76	\pm	11	$ &	1.8	& $	77	\pm	10	$ &\multicolumn{2}{c}{38.00$^1$}	&		&	25.88	& $	0.02	\pm	0.37	$ \\	
36	&	5-5	&	0.5	& $	120	\pm	11	$ &	0.8	& $	95	\pm	10	$ &		&		&		&	26.08	& $	-0.42	\pm	0.28	$ \\	
\hline
38	&	6-4	&	2.5	& $	92	\pm	11	$ &		& $				$ &	39.42	&	38.20		&		&	25.97	& $	<-2.01	\pm	0.64	$ \\	
\hline
39	&	4-1	&	1.9	& $	84	\pm	11	$ &	1.7	& $	36	\pm	10	$ &\multicolumn{2}{c}{38.43$^1$}	&	2.47	&	25.93	& $	-1.52	\pm	0.57	$ \\	
39	&	4-4	&	1.1	& $	46	\pm	11	$ &	0.03	& $	282	\pm	10	$ &		&		&		&	25.67	& $	3.26	\pm	0.45	$ \\	
\hline
40	&	4-1	&	1.4	& $	1215	\pm	25	$ &	1.0	& $	487	\pm	10	$ &	39.18	&	37.68	&	6.48	&	27.09	& $				$ \\
40	&	4-2	&	2.5	& $	493	\pm	17	$ &	2.3	& $	504	\pm	10	$ &		&		&		&	26.70	& $				$ \\	\cline{4-4} \cline{6-6}
40	&	4-Total	&		& $	1708	\pm	30	$ &		& $	991	\pm	14	$ &		&		&		&	27.24	& $	-0.98	\pm	0.13	$ \\	
\hline
41	&	4A-2	&	2.3	& $	48	\pm	11	$ &		& $				$ &	38.58	&	38.56	&	1.96	&	25.68	& $	<-0.84	\pm	0.73	$ \\	
\hline
43	&	4A-5	&	2.1	& $	33	\pm	11	$ &	2.16	& $	48	\pm	10	$ &\multicolumn{2}{c}{38.77$^1$}	&	2.2	&	25.52	& $	0.67	\pm	0.72	$ \\	
\enddata
\tablenotetext{1}{ The luminosity for these sources is in the 0.1-10.0
keV band (from Table~1).}
\end{deluxetable}

\clearpage

% FIGURE CAPTIONS

\begin{figure}
\caption{ Adaptivelly smoothed images of the Antennae in the  soft
(0.3-2.0keV) and medium (2.0-4.0keV)  bands. 
The discrete sources are represented by their  $3\sigma$ source ellipses.
 The numbering convention is the same as in Table~1 of Paper II. Light
blue ellipses indicate sources with an extended component, green
ellipses indicate variable sources and black ellipses indicate highly
obscured sources.}
\end{figure}

\begin{figure}
\caption{The composite spectra of the individual sources in the
Antennae in the three different luminosity bins described in the text. In each
figure the top panel shows the spectrum with the best fit model and
the bottom panel shows the residuals after the fit.  In the top left
figure the dotted line shows the disk-BB model, the dashed line shows
the PO component and the dash-dot line shows the RS component. In the
top right figure the dashed line shows the disk-BB component, and the
dotted and dash-dot line show the two RS components. In the bottom
figure the dashed line shows the PO component and the dash-dot line
shows the RS component. }
\end{figure}

\begin{figure}
\caption{HST-WFPC2 H$\alpha$ (a; top left), U (b; top right) and I (c;
bottom
left) band images of the Antennae  showing the
X-ray sources together with the optical sources. The arrow in the top
left corner points towards the North. Yellow ellipses
correspond to the $3\sigma$ positional ellipses of
X-ray sources with luminosities above $10^{39}$\ergs, green ellipses
 to sources in the  $3\times10^{38}-10^{39}$\ergs~ range and blue ellipses
to sources below  $3\times10^{38}$\ergs. The red symbols mark the
bright optical sources from Whitmore \etal (1999) (X's for young
clusters, crosses for intermediate age clusters, diamonds for globular
clusters and circles for foreground stars). (d) In the top panel
we plot the offsets between the X-ray source and the most nearby
optical counterpart from the papers of Whitmore \etal (1999) and
Whitmore \& Schweizer (1995) for the sources with luminosities above
$10^{39}$\ergs~ (left) and between $3\times10^{38}-10^{39}$\ergs~
(right). The vectors are proportional to the magnitude of the offset.
The numbers give the sources ID. In the bottom panel we plot the
offsets of the X-ray sources from the I-band optical sources detected
down to a 7$\sigma$ level (see text for details). The vectors for
sources 12 and 14 extentend beyond the left side of the image. The
offsets of sources 13 and 14 are in the same direction.}
\end{figure}

\begin{figure}
\caption{ Soft (left) and medium (right) band  X-ray images of the
Antennae with overlaid the X-ray and radio sources. Yellow, green and
blue  ellipses show the $3\sigma$ positional ellipses for X-ray
sources with luminosities above $10^{39}$\ergs, between
$3\times10^{38}$ and $10^{39}$\ergs, and below $3\times10^{38}$\ergs,
respectively. The red symbols correspond to the radio sources detected
by Neff \& Ulvestand (2000) (crosses and X's mark flat and steep radio
sources respectively). In the image North is up and East is left.}
\end{figure}

\begin{figure}
\caption{ A VLA HI map of the Antennae  (from Hibbard \etal 2001) 
with the position of the X-ray
sources marked by their $3\sigma$ ellipses. The contours correspond
to column densities of (17.5, 87.5, 175.5,
262.5)$\times10^{22}\rm{cm^{-2}}$. (North is up and East is left).}
\end{figure}

\setcounter{figure}{0}

\begin{figure}
\rotatebox{0}{\includegraphics[height=9.0cm]  {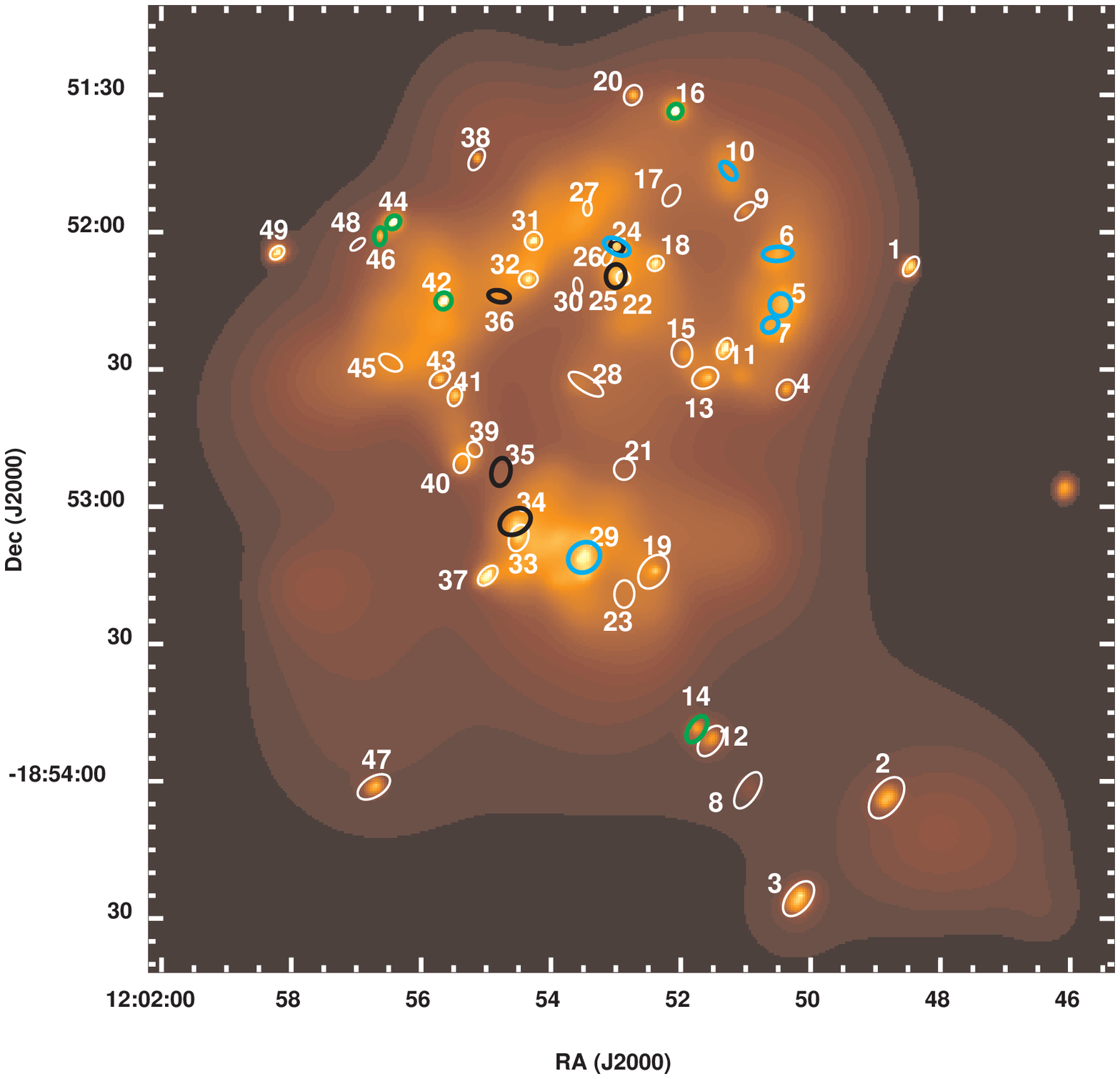}} 
\rotatebox{0}{\includegraphics[height=9.0cm]  {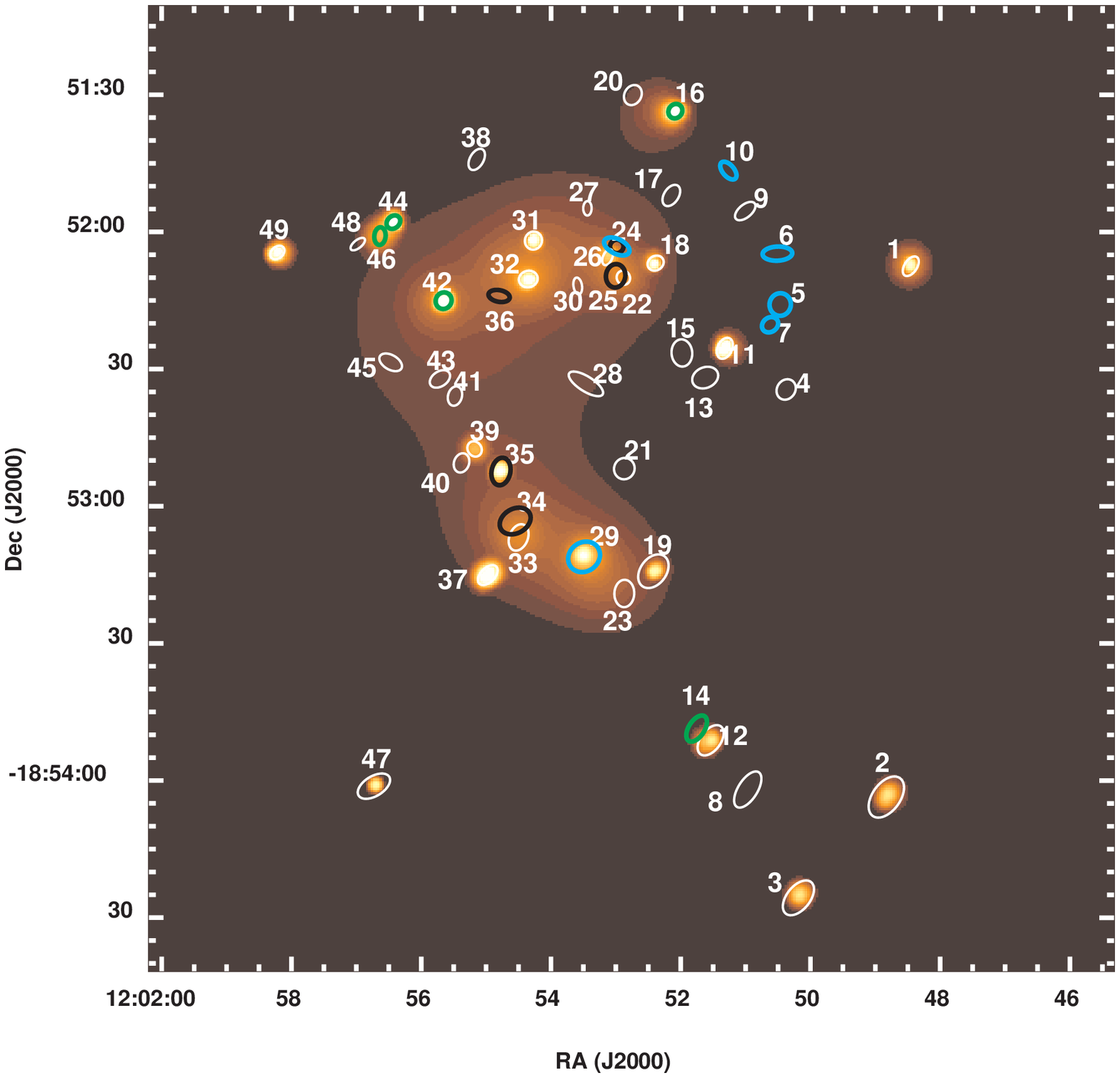}} 
\caption{}
\end{figure}		

\begin{figure}				                
\rotatebox{270}{\includegraphics[height=7.0cm]{f2a.eps}} 
\rotatebox{270}{\includegraphics[height=7.0cm]{f2b.eps}} 
\rotatebox{270}{\includegraphics[height=7.0cm]{f2c.eps}} 
\caption{}
\end{figure}	

\begin{figure}				                
%\rotatebox{360}{\includegraphics[width=18.0cm]{f3.eps}}
\caption{}
\end{figure}				                
					                
\begin{figure}				                
%\rotatebox{360}{\includegraphics[height=9.0cm]{f4.eps}}
\caption{}
\end{figure}				      
					                
\begin{figure}				                
%\rotatebox{360}{\includegraphics[height=9.0cm]{f5.eps}}          
\caption{}
\end{figure}


\begin{thebibliography}{}
\bibitem[Aretxaga et al.(2001)]{2001MNRAS.325..636A} Aretxaga, I., 
Terlevich, E., Terlevich, R.~J., Cotter, G., \& D{\' i}az, {\' A}ngeles I.\ 
2001, \mnras, 325, 636 

\bibitem[Arnett 1996]{}Arnett D., 1996, Supernonae and nucleosynthesis,
Princeton University Press

%\bibitem[Bin Mer]{}Binney J. \& Merrifield M., 1998, Galactic Astronomy, 
%Princeton University Press

\bibitem[Biretta 2000]{}Biretta, J.  \etal 2000, WFPC2 Instrument
Handbook, Version 5.0 (Baltimore, STScI) 

\bibitem[Condon \& Yin(1990)]{1990ApJ...357...97C} Condon, J.~J.~\& Yin, 
Q.~F.\ 1990, \apj, 357, 97 

%\bibitem[Cordes \& Chernoff(1998)]{1998ApJ...505..315C} Cordes, J.~M.~\& 
%Chernoff, D.~F.\ 1998, \apj, 505, 315 

\bibitem[Esin, McClintock, \& Narayan(1997)]{1997ApJ...489..865E} Esin, 
A.~A., McClintock, J.~E., \& Narayan, R.\ 1997, \apj, 489, 865 

\bibitem[Fabian 1996]{} Fabian A., \& Terlevich R., 1996  MNRAS,  280, 5

\bibitem[Fabbiano 1989]{f89} Fabbiano, G. 1989, Ann. Rev. Ast. Ap., 27, 87

\bibitem[Fabbiano 1995]{f95} Fabbiano, G. 1995, in X-ray Binaries,
ed.\ W. H. G. Lewin, J. van Paradijs, \& E. P. J. van den Heuvel
(Cambridge: University Press), p.\ 390

\bibitem[Fabbiano \etal\ 1997]{}Fabbiano, G., Schweizer, F., \& Mackie, G.
1997, \apj, 478, 542

\bibitem[Fabbiano \etal\ 2001]{} Fabbiano, G., Zezas, A., \& Murray, S.
2001, \apj, 554, 1035 (Paper~I)

%\bibitem[Filipenko 1989]{} Filippenko, A., 1989, \aj  97, 726

\bibitem[Fischer et al.(1996)]{1996A&A...315L..97F} Fischer, J.~et
al., 1996, A\&A, 315, 97 

\bibitem[Franco et al.(1993)]{1993RMxAA..27..133F} Franco, J., Miller, W., 
Cox, D., Terlevich, R., Rozyczka, M., \& Tenorio-Tagle, G.\ 1993, Revista 
Mexicana de Astronomia y Astrofisica, vol.~27, 133 

%\bibitem[Frank et al.(1992)]{} Frank , King A. \& Raine J., Accretion
%Power in Astrophysics, Cambridge University Press

%\bibitem[Freeman etal 2001]{} Freeman P.E., Kashyap V., Rosner R. \&
%Lamb D.Q., 2001a, astro-ph/0108429

%\bibitem[Freeman etal 2001]{} Freeman P.E., Doe S. \& Siemiginowska
%A., 2001b, astro-ph/0108426

%\bibitem[Fryer \& Kalogera(2001)]{2001ApJ...554..548F} Fryer, C.~L.~\& 
%Kalogera, V.\ 2001, \apj, 554, 548 

%\bibitem[Gao, Lo, Lee, \& Lee(2001)]{2001ApJ...548..172G} Gao, Y., Lo, 
%K.~Y., Lee, S.-W., \& Lee, T.-H.\ 2001, \apj, 548, 172 

%\bibitem[Garmire 1997]{}Garmire, G. P. 1997, AAS, 190, 3404

%\bibitem[Gehrels(1986)]{1986ApJ...303..336G} Gehrels, N.\ 1986, \apj, 303, 
%336 

\bibitem[Georg 2001]{} Georgakakis A., Hopkins  A. M., Caulton A.,
Wiklind T.,  Terlevich A. I. \&  Forbes D. A., 2001, astro-ph/0105435

\bibitem[Gilbert et al.(2000)]{2000ApJ...533L..57G} Gilbert, A.~M.~et al.\ 
2000, \apjl, 533, L57 

\bibitem[Gordon, Koribalski, \& Jones(2001)]{2001MNRAS.326..578G} Gordon, 
S., Koribalski, B., \& Jones, K.\ 2001, \mnras, 326, 578 

\bibitem[Hibb 2001]{} Hibbard, J. E., Van der Hulst, J. M., Barnes
J. E., Rich R. M., 2001, astro-ph/0110581

%\bibitem[Kilgard et al.(2001)]{2001MNRAS.321L..29K} Kilgard, R.,
%Kaaret, P., Krauss, M.,  Prestwich,  A.~H., Raley, M. \& Zezas, A.,
%2001, submitted to \apj

\bibitem[King et al.(2001)]{2001ApJ...552L.109K} King, A.~R., Davies, 
M.~B., Ward, M.~J., Fabbiano, G., \& Elvis, M.\ 2001, \apjl, 552, L109 

\bibitem[Kubota et al.(2001)]{2001ApJ...547L.119K} Kubota, A., Mizuno, T., 
Makishima, K., Fukazawa, Y., Kotoku, J., Ohnishi, T., \& Tashiro, M.\ 2001, 
\apjl, 547, L119 

\bibitem[Kunze et al.(1996)]{1996A&A...315L.101K} Kunze, D.~et al.\ 1996, 
\aap, 315, L101 

\bibitem[Leitherer et al.(1999)]{1999ApJS..123....3L} Leitherer, C.~et al.\ 
1999, \apjs, 123, 3 

\bibitem[Makishima \etal\ 2000]{}Makishima, K. \etal\ 2000, \apj, 535, 632

\bibitem[Meurs \& van den Heuvel(1989)]{1989A&A...226...88M} Meurs, 
E.~J.~A.~\& van den Heuvel, E.~P.~J.\ 1989, \aap, 226, 88 

\bibitem[Mengel et al.(2001)]{2001ApJ...550..280M} Mengel, S., Lehnert, 
M.~D., Thatte, N., Tacconi-Garman, L.~E., \& Genzel, R.\ 2001, \apj, 550, 
280 

\bibitem[Mihos \& Hernquist(1996)]{1996ApJ...464..641M} Mihos, J.~C.~\& Hernquist, L.\ 1996, \apj, 464, 641 

\bibitem[Miller et al.(2001)]{2001ApJ...546.1055M} Miller, J.~M., Fox, 
D.~W., Di Matteo, T., Wijnands, R., Belloni, T., Pooley, D., Kouveliotou, 
C., \& Lewin, W.~H.~G.\ 2001, \apj, 546, 1055 

\bibitem[Mirabel et al.(1998)]{1998A&A...333L...1M} Mirabel, I.~F.~et al.\ 
1998, \aap, 333, L1 

\bibitem[Moran, Lehnert, \& Helfand(1999)]{1999ApJ...526..649M} Moran, 
E.~C., Lehnert, M.~D., \& Helfand, D.~J.\ 1999, \apj, 526, 649 

\bibitem[Nagase(1989)]{1989PASJ...41....1N} Nagase, F.\ 1989, \pasj, 41, 1 

\bibitem[Neff \& Ulvestad(2000)]{2000AJ....120..670N} Neff, S.~G.~\& 
Ulvestad, J.~S.\ 2000, \aj, 120, 670 

\bibitem[Nowak(1995)]{1995PASP..107.1207N} Nowak, M.~A.\ 1995, \pasp, 107, 
1207 

\bibitem[Plewa 1995]{pl95} Plewa T.,   1995, \mnras, 275, 143

%\bibitem[Portegies Zwart, Makino, McMillan, \& 
%Hut(1999)]{1999A&A...348..117P} Portegies Zwart, S.~F., Makino, J., 
%McMillan, S.~L.~W., \& Hut, P.\ 1999, \aap, 348, 117 

\bibitem[Ptak \& Griffiths(1999)]{1999ApJ...517L..85P} Ptak, A.~\& Griffiths, R.\ 1999, \apjl, 517, L85

\bibitem[Roberts 2000]{} Roberts T., \& Warwick R., 2000, MNRAS, 315, 98

\bibitem[Schlegel 1995]{}Schlegel E., 1995, Reports of Progress in
Physics, 58, 1375

\bibitem[Strickland, Ponman, \& Stevens(1997)]{1997A&A...320..378S} Strickland, D.~K., Ponman, T.~J., \& Stevens, I.~R.\ 1997, \aap, 320, 378   

\bibitem[Strickland, Heckman, Weaver, \& Dahlem(2000)]{2000AJ....120.2965S} Strickland, D.~K., Heckman, T.~M., Weaver, K.~A., \& Dahlem, M.\ 2000, \aj, 120, 2965

\bibitem[Strickland \& Stevens(2000)]{2000MNRAS.314..511S} Strickland, D.~K.~\& Stevens, I.~R.\ 2000, \mnras, 314, 511

\bibitem[str 01]{} Strickland, D.K., Colbert,   E.J.M.,  Heckman,
T.M., Weaver, K.A., Dahlem, M.,  Stevens, I.R., 2001, astro-ph/0107115

\bibitem[Tan Lew 1995]{f95} Tanaka, F. \& Lewin W. 1995, in X-ray Binaries,
ed.\ W. H. G. Lewin, J. van Paradijs, \& E. P. J. van den Heuvel
(Cambridge: University Press), p.\ 126

\bibitem[Tanaka \& Shibazaki(1996)]{} Tanaka, Y.~\& 
Shibazaki, N.\ 1996, \araa, 34, 607 

\bibitem[Terl 1994]{t95} Terlevich, R. 1994, in Circumstellar Media in
the Late Stages of Stellar Evolution,
ed.\ R.E.S. Clegg, I.R. Stevens,  W.P.S Meikle, J. van Paradijs, 
(Cambridge: University Press), p.\ 153

\bibitem[Toomre \& Toomre(1972)]{1972ApJ...178..623T} Toomre, A.~\& Toomre, J.\ 1972, \apj, 178, 623

\bibitem[van den Heuvel, Portegies Zwart, \& 
Kaper(2000)]{2000A&A...364..563V} Van den Heuvel, E.~P.~J., Portegies 
Zwart, S.~F.~B.~D., \& Kaper, L.\ 2000, \aap, 364, 563 

\bibitem[van Paradijs 1999]{} Van Paradijs, 
J., 1999, in the Many Faces of Neutron Stars, ed.\ R. Buccheri, J. van Paradijs, M.A.
     Alpar (Kluwer Academic Publishers), p.\ 279  (astro-ph/9802177)

\bibitem[van Paradijs \& White(1995)]{1995ApJ...447L..33V} Van Paradijs, 
J.~\& White, N.\ 1995, \apjl, 447, L33 

\bibitem[Van Speybroeck \etal\ 1997]{} Van Speybroeck, L., Jerius D., Edgar, R. J., Gaetz, T. J., Zhao, P. \& Reid, P. B.1997, Proc. SPIE 3113, 89 

\bibitem[Vigroux et al.(1996)]{1996A&A...315L..93V} Vigroux, L.~et al.\ 
1996, \aap, 315, L93 

\bibitem[Watarai, Fukue, Takeuchi, \& Mineshige(2000)]{2000PASJ...52..133W} 
Watarai, K., Fukue, J., Takeuchi, M., \& Mineshige, S.\ 2000, \pasj, 52, 133 

\bibitem[Watson 1990]{} 
Watson M. G., 1990, in Windows in Galaxies, eds, Fabbiano G., Gallagher J.S., and Renzini A. 

\bibitem[Weisskopf \etal\ 2000]{}Weisskopf, M., Tananbaum, H., Van Speybroeck, L. \& O'Dell, S.
2000, Proc. SPIE 4012 (astro-ph 0004127)

\bibitem[White \& van Paradijs(1996)]{1996ApJ...473L..25W} White, N.~E.~\& van Paradijs, J.\ 1996, \apjl, 473, L25

\bibitem[Whitmore \& Heyer(1995)]{}Whitmore \& Heyer, 1995, http://www.stsci.edu/instrument-news/isr/wfpc2/9504/9504\_1.html
\bibitem[Whitmore \& Schweizer(1995)]{1995AJ....109..960W} Whitmore, 
B.~C.~\& Schweizer, F.\ 1995, \aj, 109, 960 

\bibitem[Whitmore et al.(1999)]{1999AJ....118.1551W} Whitmore, B.~C., 
Zhang, Q., Leitherer, C., Fall, S.~M., Schweizer, F., \& Miller, B.~W.\ 
1999, \aj, 118, 1551 

\bibitem[Wilson, Scoville, Madden, \& 
Charmandaris(2000)]{2000ApJ...542..120W} Wilson, C.~D., Scoville, N., 
Madden, S.~C., \& Charmandaris, V.\ 2000, \apj, 542, 120 

\bibitem[Zezas \etal\ 1999]{} Zezas, A., Georgantopoulos, I. \& Ward, M., 1999, MNRAS, 308, 302

\bibitem[Zezas \etal\ 2001]{}Zezas A., Fabbiano G.,  Prestwich A., Ward M., \& Murray S., 
2001,  Procs of The Central Kiloparsec of Starbursts and 
AGN: The La Palma Connection, ASP Conf Series, Vol. 249, 425, Eds. J.H. 
Knapen, J.E. Beckman, I. Shlosman, and T.J. Mahoney
(astro-ph/0109302)

\bibitem[Zezas \etal\ 2002]{} Zezas, A. \& Fabbiano G., 2002, submitted to \apj (Paper~IV)

\bibitem[Zezas \etal\ 2002]{} Zezas, A. \& Fabbiano G., Rots A.H. \&
Murray S.S.,  2002, submitted to \apj (Paper~III)

\bibitem[Zhang, Fall, \& Whitmore(2001)]{2001ApJ...561..727Z} Zhang,
Q., Fall, S.~M., \& Whitmore, B.~C.\ 2001, \apj, 561, 727 
\end{thebibliography}
\end{document}